\documentclass[a4paper,11pt,DIV=12,draft=false]{scrartcl}

\usepackage[ttscale=0.9]{libertine}
\usepackage[T1]{fontenc}
\usepackage[utf8]{inputenc}

\usepackage{graphicx}
\usepackage{nicefrac}
\usepackage{caption}
\usepackage{subcaption}
\usepackage{booktabs}
\usepackage{url}
\usepackage{mathrsfs}
\usepackage{amssymb}
\usepackage{amsmath}
\usepackage{scrlayer-scrpage}
\usepackage[affil-it]{authblk}
\usepackage[numbers,sort&compress]{natbib}
\usepackage[titletoc,title]{appendix}
\usepackage[protrusion=true,expansion,kerning=true,tracking=true,final]{microtype}
\usepackage[
            colorlinks=true,
            linkcolor=hblue,
            citecolor=hgreen,
            filecolor=hblue,
            urlcolor=hred
            ]{hyperref}
\usepackage{enumitem}

\usepackage[dvipsnames,x11names]{xcolor}
\definecolor{hgreen}{rgb}{0,.3,0}
\definecolor{hred}{rgb}{.3,0,0}
\definecolor{orange}{rgb}{1,0.5,0}
\definecolor{hblue}{rgb}{0,0,.3}
\definecolor{LightGray}{gray}{0.95}
\definecolor{gray}{gray}{0.6}

\allowdisplaybreaks
\setcapindent{1em}
\setkomafont{captionlabel}{\bfseries}
\setkomafont{caption}{\itshape}
\setkomafont{titlehead}{\normalsize\sffamily}
\addtokomafont{title}{\boldmath}
\addtokomafont{section}{\boldmath}
\KOMAoptions{headinclude=false,
             footinclude=false,
             parskip=false,
             draft=false,
             overfullrule=false,
             abstract=true,
             numbers=noenddot,
             titlepage=false,
             twocolumn=false,
             twoside=false,
             DIV=12}

\usepackage[shortcuts]{extdash}

\usepackage{placeins}
\usepackage{dsfont}
\usepackage{soul}

\makeatletter
\g@addto@macro\bfseries\boldmath
\makeatother

\makeatletter
\DeclareOldFontCommand{\rm}{\normalfont\rmfamily}{\mathrm}
\DeclareOldFontCommand{\sf}{\normalfont\sffamily}{\mathsf}
\DeclareOldFontCommand{\tt}{\normalfont\ttfamily}{\mathtt}
\DeclareOldFontCommand{\bf}{\normalfont\bfseries}{\mathbf}
\DeclareOldFontCommand{\it}{\normalfont\itshape}{\mathit}
\DeclareOldFontCommand{\sl}{\normalfont\slshape}{\@nomath\sl}
\DeclareOldFontCommand{\sc}{\normalfont\scshape}{\@nomath\sc}
\makeatother

\usepackage{xspace} 

\newcommand{\pt}{\ensuremath{p_{\mathrm{T}}}\xspace}
\newcommand{\myEta}{\ensuremath{\eta}\xspace}
\newcommand{\noise}{\ensuremath{N}\xspace}
\newcommand{\viA}{\ensuremath{v^\mathrm{A}_i}\xspace}
\newcommand{\vOneA}{\ensuremath{v^\mathrm{A}_1}\xspace}
\newcommand{\vTwoA}{\ensuremath{v^\mathrm{A}_2}\xspace}
\newcommand{\viB}{\ensuremath{v^\mathrm{B}_i}\xspace}
\newcommand{\vOneB}{\ensuremath{v^\mathrm{B}_1}\xspace}
\newcommand{\vTwoB}{\ensuremath{v^\mathrm{B}_2}\xspace}

\providecommand{\keywords}[1]
{
  \small	
  \textbf{\textit{Keywords---}} #1
}

\begin{document}
\title{One flow to correct them all: \\ improving simulations in high-energy physics with a single normalising flow and a switch}

\author[a]{Caio~Cesar~Daumann%
\thanks{Corresponding author: \texttt{caio.daumann@rwth-aachen.de}}}

\author[b]{Mauro~Donega%
\thanks{\texttt{mdonega@phys.ethz.ch}}}

\author[a]{Johannes~Erdmann%
\thanks{\texttt{johannes.erdmann@physik.rwth-aachen.de}}}

\author[b]{Massimiliano~Galli%
\thanks{\texttt{masgalli@phys.ethz.ch}}}

\author[a]{Jan Lukas Sp\"ah%
\thanks{\texttt{janlukas.spaeh@physik.rwth-aachen.de}}}

\author[b]{Davide~Valsecchi%
\thanks{\texttt{dvalsecchi@ethz.ch}}}

\date{}

\affil[a]{{\large RWTH Aachen University, III. Physikalisches Institut A, Aachen, Germany}}
\affil[b]{{\large ETH Z\"urich, Z\"urich, Switzerland}}

\maketitle

\begin{abstract}
\normalsize
Simulated events are key ingredients in almost all high-energy physics analyses.
However, imperfections in the simulation can lead to sizeable differences between the observed data and simulated events.
The effects of such mismodelling on relevant observables must be corrected either effectively via scale factors, with weights
or by modifying the distributions of the observables and their correlations.
We introduce a correction method that transforms one multidimensional distribution (simulation) into another one (data) using a simple architecture based on a single normalising flow with a boolean condition.
We demonstrate the effectiveness of the method on a physics-inspired toy dataset with non-trivial mismodelling of several observables and their correlations.
\end{abstract}

\begin{center}
\end{center}

\newpage

\section{Introduction\label{sec:introduction}}

Monte Carlo simulations play a key role in the data analysis of high-energy physics experiments.
The simulation of final-state particles from scattering events and the simulation of their interactions with the detector material are used in many applications.
Important examples are the development of particle reconstruction algorithms, the calibration of the properties of the reconstructed particles, the optimization of event-level signal-background classifiers, and the estimate of signal and background contributions to selected phase spaces.
While these simulations often provide a very good description of the data, residual imperfections in the simulations can lead to sizeable deviations from the observed data.
This can result in reduced performance of algorithms that were developed based on simulated events, biased signal or background estimates, or biased calibrations.

The effect of such mismodelling on a relevant observable, for example, the efficiency for reconstructing a certain particle, is often mitigated by so-called ``scale factors'' to the simulations.
These scale factors do not attempt to correct the underlying mismodelling itself but aim to effectively remove the data-simulation differences for the given application.
They come with their associated statistical and systematic uncertainties, which can limit the sensitivity of measurements and searches in high-energy experiments.

An alternative is to address the underlying mismodelling of the relevant observables, such as the features that are used to reconstruct a certain particle.
Such an approach promises to be more general, as any algorithm that is based on the corrected features is expected to show a better agreement of simulation and data.
One approach is to derive weights for simulated events. These can be obtained from a classifier that is trained to distinguish simulation and data~\textcolor{black}{\cite{Cranmer:2015bka, Rogozhnikov:2016bdp, Andreassen:2019nnm, Diefenbacher_2020_DCTRGAN}}.
However, a disadvantage of the weighting approach is that observables that are not part of the training may show a worse agreement between simulations and data after the weights are applied.
The weighting approach may also lead to increased uncertainties due to the limited number of samples in the simulation in case of large weights.
Another approach is to morph the observables to correspond to the multidimensional target distribution so that only certain observables are modified.
This approach has been implemented with chained quantile regressions~\cite{CQR_Higgs_full_run2}, generative adversarial networks~\cite{Erdmann:2018kuh}, input convex neural networks~\cite{amos2017input} for optimal transport properties~\cite{Pollard:2021fqv}, generative diffusion networks~\cite{Butter:2023ira} and normalising flows~\cite{Algren:2023qnb, CQRwithFlows, flow4flows}.
In the latter approach, normalising flows~\cite{Tabak:2013cnz, papamakarios2021normalizing_review} are trained as a bijective transformation between the complex multidimensional distribution in the input space and a simple ``base distribution'' of the same dimensionality, often a multivariate Gaussian distribution.
Since monotonically increasing bijective transformations are used, normalising flows preserve the quantiles of the distributions, which makes them suitable for morphing.

We propose a morphing procedure based on a single normalising flow that is conditioned on a boolean (``IsData'') that encodes whether the input is drawn from the simulation or the data.
We train the normalising flow simultaneously on both datasets to learn a conditional mapping between the input distributions and the base distribution.
After the normalising flow is trained, we then map samples from simulation
to the base distribution of the flow and switch the boolean condition before mapping back to the input space, effectively considering them samples from data.
Our approach differs from previous work on normalising flows for morphing simulations.
One proposed strategy is to learn a mapping from input to base distribution only for the data and sample from the base distribution to produce corrected values for the simulation~\cite{Algren:2023qnb}.
Another strategy is a chain of normalising flows, one for each dimension in the input space, similar to chained quantile regression~\cite{CQRwithFlows}.
In Ref.~\cite{flow4flows}, five different methods were studied: The ``base transfer'' method uses a combination of two normalising flows that map the two input spaces to the same base distribution.
In the ``unidirectional transfer'' method, the base distribution for one of the normalising flows is given by the distribution in the other input space, which is then mapped to a base distribution by a second normalising flow.
The ``flows for flows'' approach extends the unidirectional transfer method to be bidirectional, using a third normalising flow, so that both input spaces are mapped to separate base distributions.
Two additional proposals in Ref.~\cite{flow4flows} extend the flows for flows with constraints on the learned transfer maps.

Our approach is novel and simple, as we propose to train only one normalising flow that learns the mapping for data and simulations simultaneously.
Similar to the base transfer method, we train a mapping to the same base distribution for both input spaces.
However, the use of only one flow simplifies the training procedure and the time spent in optimisation.
It results in an effective sharing of parameters for the mappings of the two input spaces to the base distribution.

We study our approach in a toy example that captures several aspects of realistic applications in high-energy-physics experiments: (a) the observables follow different marginal probability density functions; (b) the observables are partially correlated; (c) the probability density functions and the structure of the correlations depend on ancillary variables; (d) the probability density functions for data and simulation differ in their shapes, their correlations, and their dependence on the ancillary variables.
We investigate to what extent our morphing approach can correct the marginal distributions and their correlations, and we test whether a multivariate classifier can still separate data from the corrected simulation.
In addition, we describe a preprocessing step that transforms discontinuous input distributions into multimodal distributions and significantly improves the quality of the corrections.

We introduce the setup of the normalising flow in Section~\ref{sec:corrections} and evaluate the performance for the case of well-known two-dimensional benchmarks for generative models (checkerboard, two moons, four circles).
In Section~\ref{sec:dataGen}, we describe the generation of the physics-inspired toy dataset.
The quality of the corrections on this dataset is discussed in Section~\ref{sec:performance}.
We present our conclusions in Section~\ref{sec:conclusions}.

\section{Correcting simulations with one normalising flow}
\label{sec:corrections}

\subsection{Normalising flows for morphing distributions}

Normalising flows are a class of generative models designed for the effective learning and sampling of multivariate probability distributions. They are constructed by parametrising a transformation that performs the mapping of a complex input distribution to a tractable base distribution of the same dimension, $d$.
Such a transformation, $f$, must be invertible and hence ensure a one-to-one correspondence between the input probability density function (PDF), $p_x(\mathbf{x})$, and the base PDF, $p_z(\mathbf{z})$, i.e.~$f: \mathbb{R}^{d} \rightarrow \mathbb{R}^{d}$.
Given that the composition of multiple invertible functions remains invertible, it is common practice to construct the transformation $f$ as a composition of $N$ invertible transformations to increase the expressiveness of the model~\cite{papamakarios2021normalizing_review, Kobyzev_2021}, ${f = f_{1} \circ \ldots \circ f_{N}}$.

If $f$ is differentiable, one can express the PDF in the input space, $p_{x}(\boldsymbol{x})$, as a function of the base PDF and the Jacobian matrix of the transformation, $J_f$, with the change of variables formula as
\begin{equation}
    p_{x}(\bold{x}) = p_{z}(f(\bold{x})) \left| \det J_{f}(\bold{x})  \right|\,.
    \label{eq:pdfs}
\end{equation}
For a composition of transformations, the log-likelihood function becomes
\begin{equation}
    \log p_{x}(\bold{x}) = \log p_{z}(f(\bold{x})) + \sum_{i=1}^{N} \log \left| \det J_{f_{i}}(y_i) \right|\,,
    \label{eq:loss2}
\end{equation}
where the $y_i = f_{i+1} \circ \dots \circ f_N$ represent the outputs of intermediate flows.
With a suitable choice of the base PDF \(p_{z}\) and the transformation $f$, equation~\eqref{eq:loss2} has a closed form and can be used to train the normalising flow with negative log-likelihood as loss.
The transformations $f_{i}$ are typically parameterised by neural networks, referred to as ``auxiliary networks'' in the following.

We used masked autoregressive flows~\cite{papamakarios2018masked_MAF} based on spline transformations~\cite{müller2019neural_linear_quadratic_splines,durkan2019cubicspline,linear_rational_spline,Spline_flows}.
The autoregressive property of the flow leads to an efficient computation of the Jacobian determinant.
The auxiliary networks are implemented as multilayer perceptrons with masked connections, for which we use MADE blocks~\cite{germain2015made} (Masked Autoencoder for Distribution Estimation).
Using splines in normalising flows has the advantage that they can approximate more complex distributions compared to affine transformations while still being simple to invert.
We use neural spline flows~\cite{Spline_flows}, which are based on monotonic rational quadratic splines.
The monotonicity of the transformation is important for morphing distributions, because they ensure that the quantiles of the distributions are preserved during the transformation.
\textcolor{black}{ When a transformation is monotonic, the order of data points is maintained. This means that the same proportion of data points will fall below any given threshold in both the original and transformed distributions. Consequently, the quantiles, which represent these thresholds, remain unchanged. }

\begin{figure}[ht]
    \centering
    \includegraphics[width=0.62\textwidth]{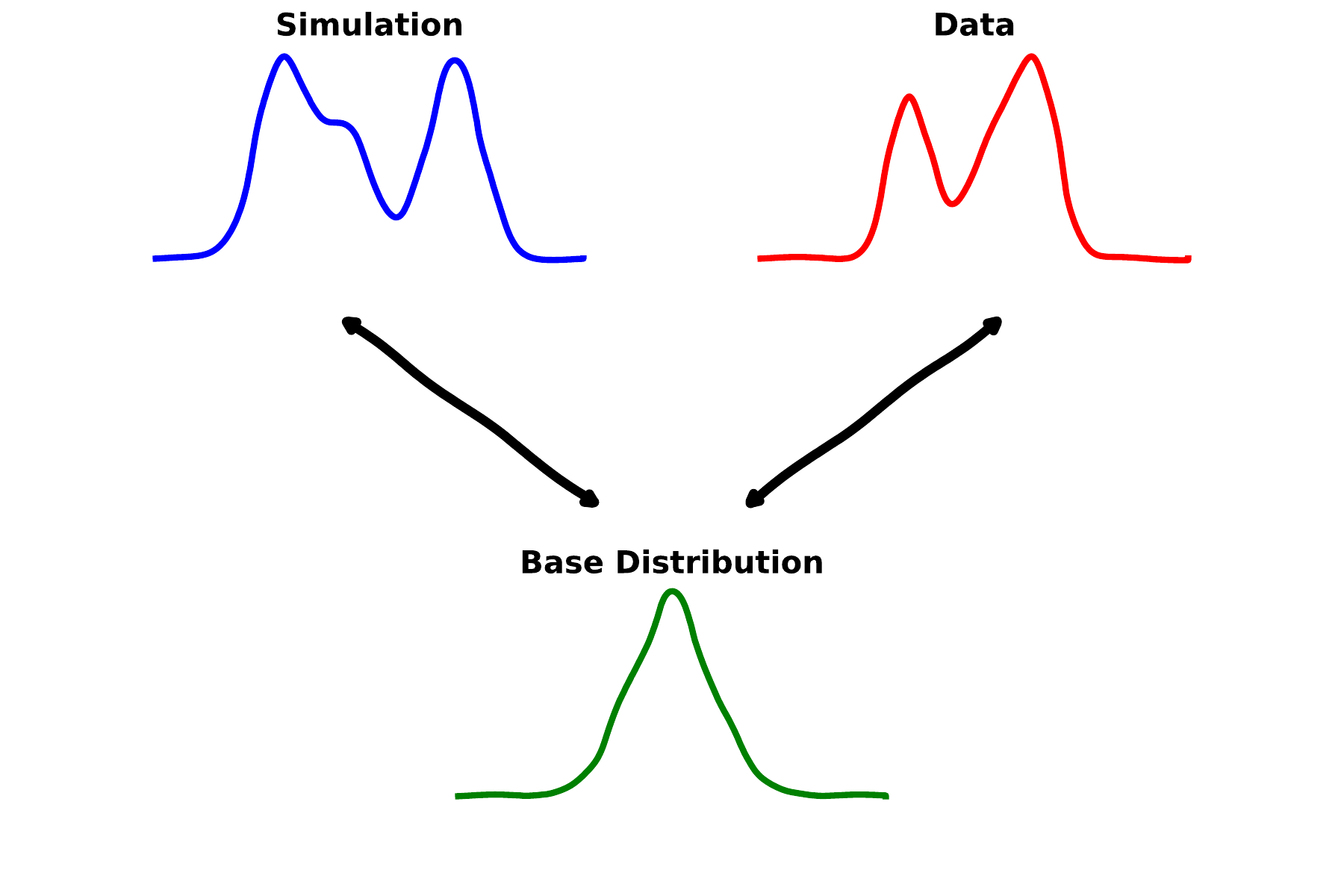}\\
    \vspace{0.5cm}
    \includegraphics[width=\textwidth]{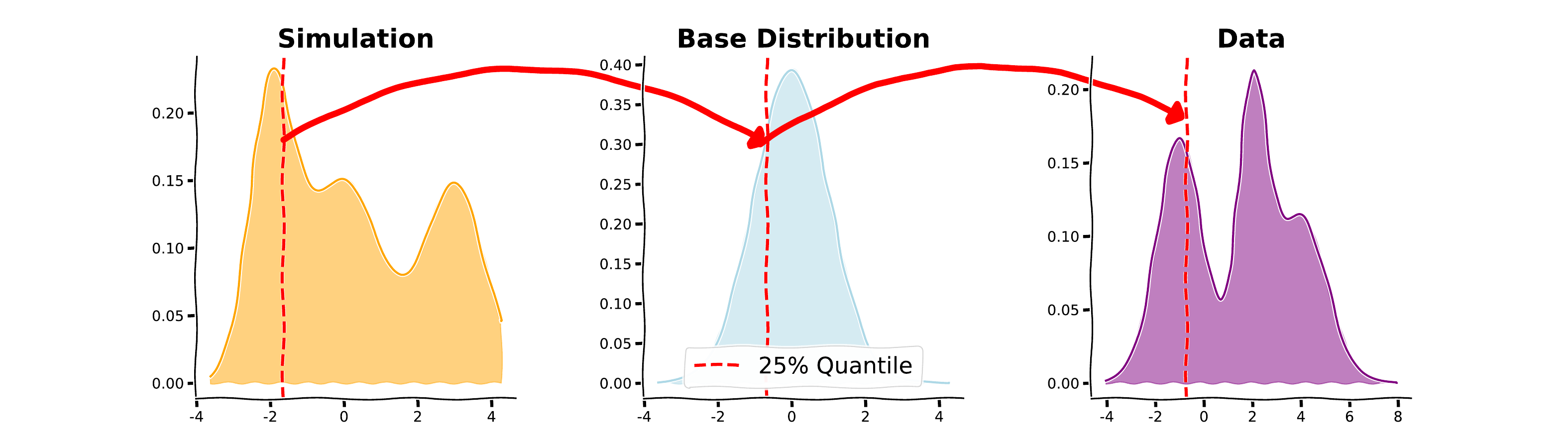}
    \caption{Top: Illustration of the single-flow morphing. The normalising flow is trained to map both data and simulation to the same base distribution.
    The flow is conditioned on a boolean that encodes whether the input is drawn from simulation or data.
    Bottom: Illustration of the preservation of quantiles during the morphing from simulation to data space using the base distribution as an intermediary.
    }    
    \label{fig:flow_schematic}
\end{figure}

The normalising flow is simultaneously trained on both datasets, i.e.~simulation and data, mapping them to a shared base distribution.
This approach is illustrated in Figure~\ref{fig:flow_schematic}~(top).
We condition the flow on a boolean variable (``IsData''), which enables the learning of distinct mappings to the common base distribution for both, simulation and data.
To correct the simulation, we use the trained flow to map samples from simulation to the base distribution, switch the boolean, and use the inverse transformation to map back to the data input space.
This procedure effectively performs a quantile morphing between simulation and data, using the base distribution as an intermediary as depicted in Figure~\ref{fig:flow_schematic}~(bottom).

\subsection{Two-dimensional benchmarks}
\label{sec:2D}

For a first evaluation of the single-flow approach, we check its capability to morph two sets of commonly used two-dimensional benchmark datasets for generative models.
In the first test, we morph the checkerboard distribution to the four-circle distribution.
In the second test, we morph the checkerboard distribution to the two-moons distribution.
In both cases, we also check the inverse morphing performance of the flow.
The checkerboard and four-circles datasets are taken from the GitHub repository of Ref.~\cite{flow4flows}, the two-moons dataset was generated using scikit-learn~\cite{sklearn}. 
One million samples were drawn from each distribution, using $60\%$ for training, $10\%$ for validation and $30\%$ for testing.

We use the neural spline implementation of the Zuko package~\cite{zuko} in PyTorch~\cite{pytorch}.
\textcolor{black}{We set the number of bins in each spline to $8$.} 
For the MADE blocks, we use two hidden layers with $128$ nodes each and ReLU activation.
We use the ADAM optimiser~\cite{Adam_optimizer} with a cosine annealing learning-rate scheduler~\cite{cossine_annealing}.
The training is performed until the validation loss does not improve over $15$ epochs.
For the two-dimensional benchmark examples, the normalising flow is composed of four monotonic rational-quadratic transformations.

\begin{figure}[h!]
    \centering
    \begin{subfigure}{0.49\textwidth}
        \includegraphics[width=\linewidth]{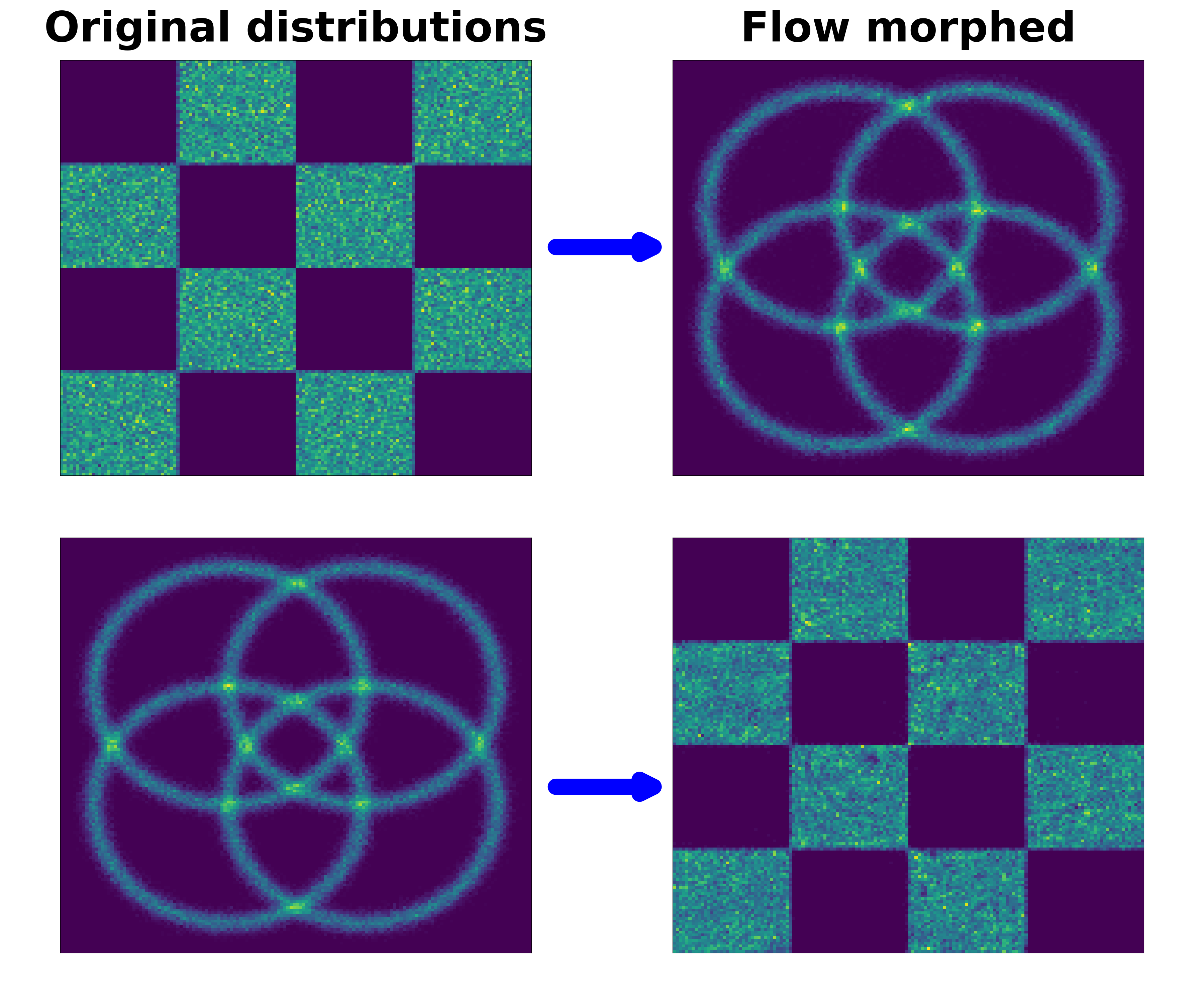}
        \label{fig:img1}
    \end{subfigure}
    \hfill
    \begin{subfigure}{0.49\textwidth}
        \includegraphics[width=\linewidth]{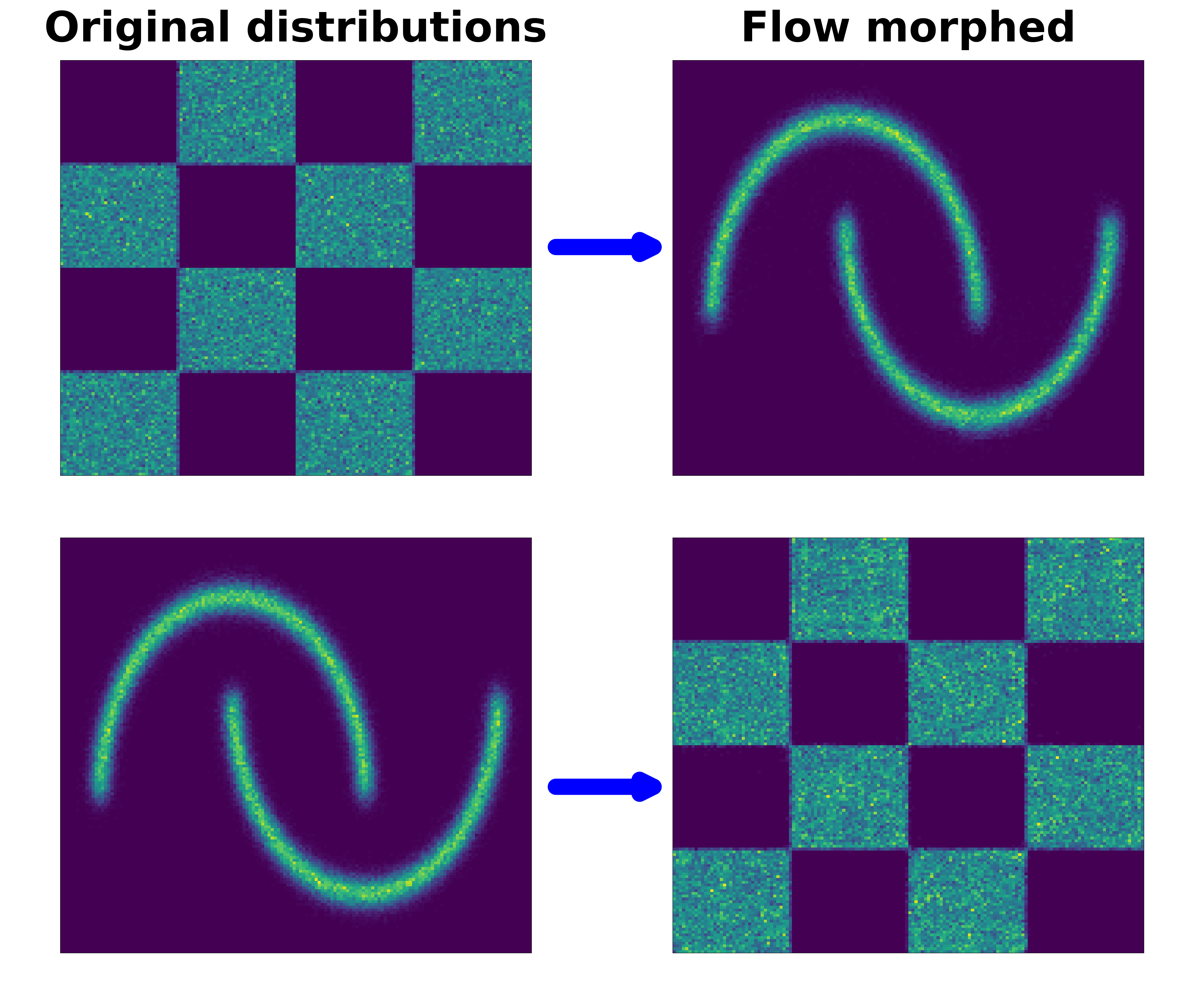}
        \label{fig:img2}
    \end{subfigure}
    \hfill
    \caption{The upper plots show the morphing from the checkerboard distribution into the four-circles distribution (left) and into the two-moons distribution (right). The lower plots illustrate the inverted transformation.
    }
    \label{fig:2d_toy}
\end{figure}

Figure~\ref{fig:2d_toy} shows the result of the morphing, where each plot displays the original distributions on the left with the resulting morphed distributions on the right.
Our approach is able to reproduce also the sharp edges and discontinuous features in both sets of distributions.
We note that it is simple to extend the boolean condition to a one-hot encoding in order to morph between more than two domains with a single flow. We illustrate this in a three-domain example in Appendix~\ref{app:3D}.

\FloatBarrier
\newpage

\section{Generation of the physics-inspired dataset}
\label{sec:dataGen}

The toy dataset is divided into the two toy classes ``data'' and ``simulation'' and includes seven variables.
Three of the variables are ancillary variables that are loosely inspired by kinematic features in typical high-energy physics applications, whereas the other four variables can be interpreted as informative features that discriminate between signal and background.
The informative features are conditioned on the ancillary variables to enhance the complexity of the multivariate distribution and make the dataset more realistic.
The structure of the conditions is different for data and simulation.
This encodes the mismodelling deep into the structure of the multivariate distribution, making the task of correcting simulated events to data more challenging.
In addition to the conditioning, non-trivial correlations are included between several of the seven variables.

The ancillary features ``\pt'' and ``\myEta'' are inspired by the typical shapes of transverse momentum and absolute pseudorapidity distributions from decays of massive particles in collider experiments.
The distribution of the \pt\ variable is exponential with different scale parameters for data and simulation.
The \myEta\ variable is drawn from a uniform distribution in the interval $[-2,2]$, and a Gaussian smearing with a mean of unity and different standard deviations for data and simulation is applied to smooth the edges of the interval.
Finally, the absolute value is taken to obtain the \myEta\ values.
The third ancillary feature, the noise ``\noise'' is uniform in the interval $[0,3]$.
The noise variable can either correspond to the azimuthal angle, which might have non-trivial correlations with informative features in the case of detector asymmetries, or to a variable that is related to the pileup conditions at hadron-collider experiments.

The four informative features are divided into two families $\mathrm{A}$ and $\mathrm{B}$.
The two variables \vOneA\ and \vTwoA\ are drawn from uncorrelated Gaussian distributions.
The mean and the standard deviation are different for data and simulation and depend on the ancillary features.
With increasing \myEta, the distributions are shifted to the left.
The distributions become increasingly narrow for high \pt, whereas larger noise leads to broader distributions.
Overall, both \viA distributions roughly resemble Gaussian distributions, although the non-trivial conditions lead to slightly non-Gaussian effects, such as heavy or asymmetric tails and flattened or compressed peaks.
The distributions for the \vOneB\ and \vTwoB\ variables
are discontinuous.
A certain fraction of the samples is assigned a value of zero and the rest of the values are drawn from shifted exponential distributions.
This corresponds to a density mixture of a Dirac delta at zero with a tail to the right.
The fraction of samples at zero increases with \pt\ and decreases with larger \noise\ values.
The scale of the exponential distributions for the events in the tail increases with larger \myEta\ and \noise\ values.
Again, these effects differ in magnitude between data and simulation.

\begin{figure}[p]
    \centering
    \begin{subfigure}{0.33\textwidth}
        \includegraphics[width=\linewidth]{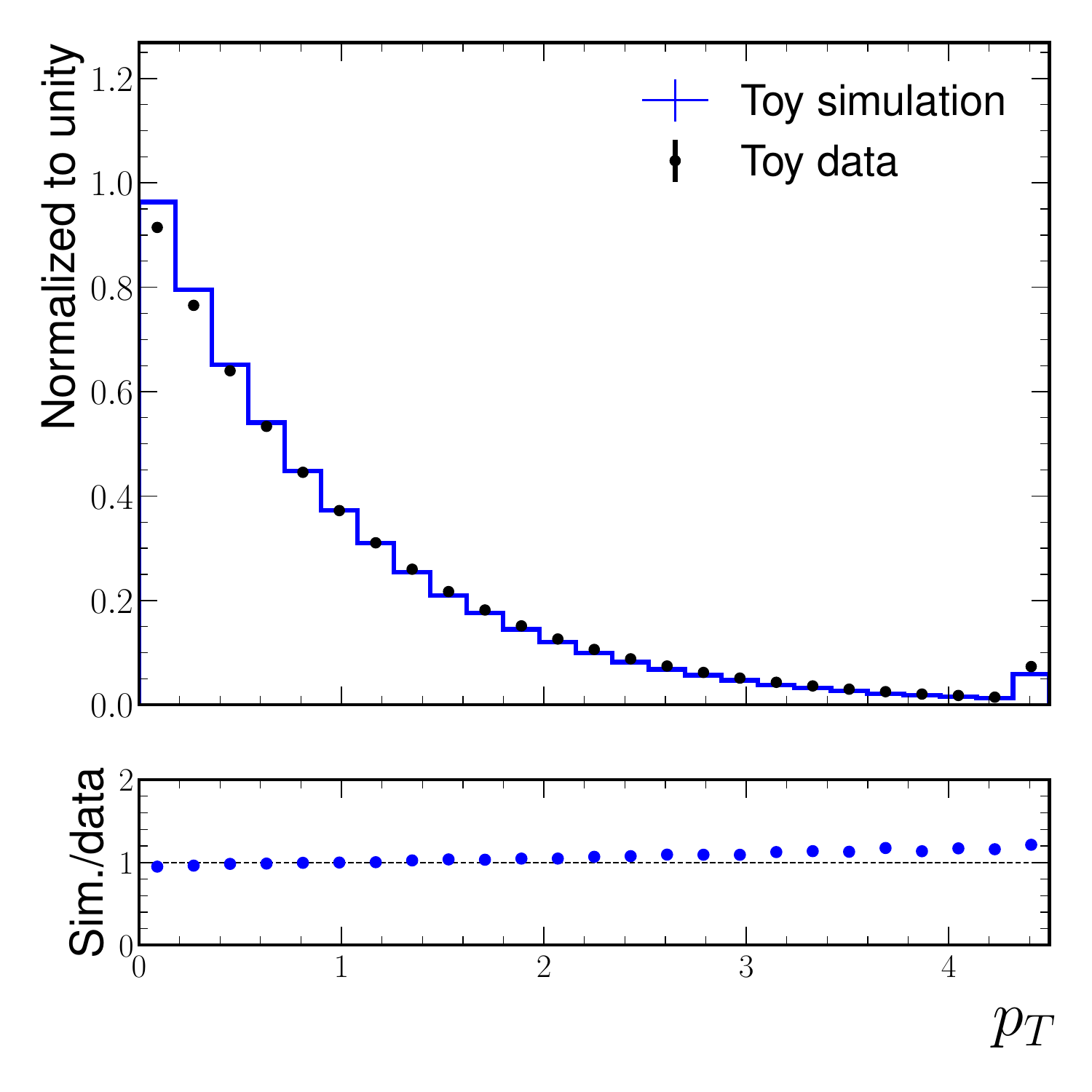}
    \end{subfigure}
    \hspace{1cm}
    \begin{subfigure}{0.33\textwidth}
        \includegraphics[width=\linewidth]{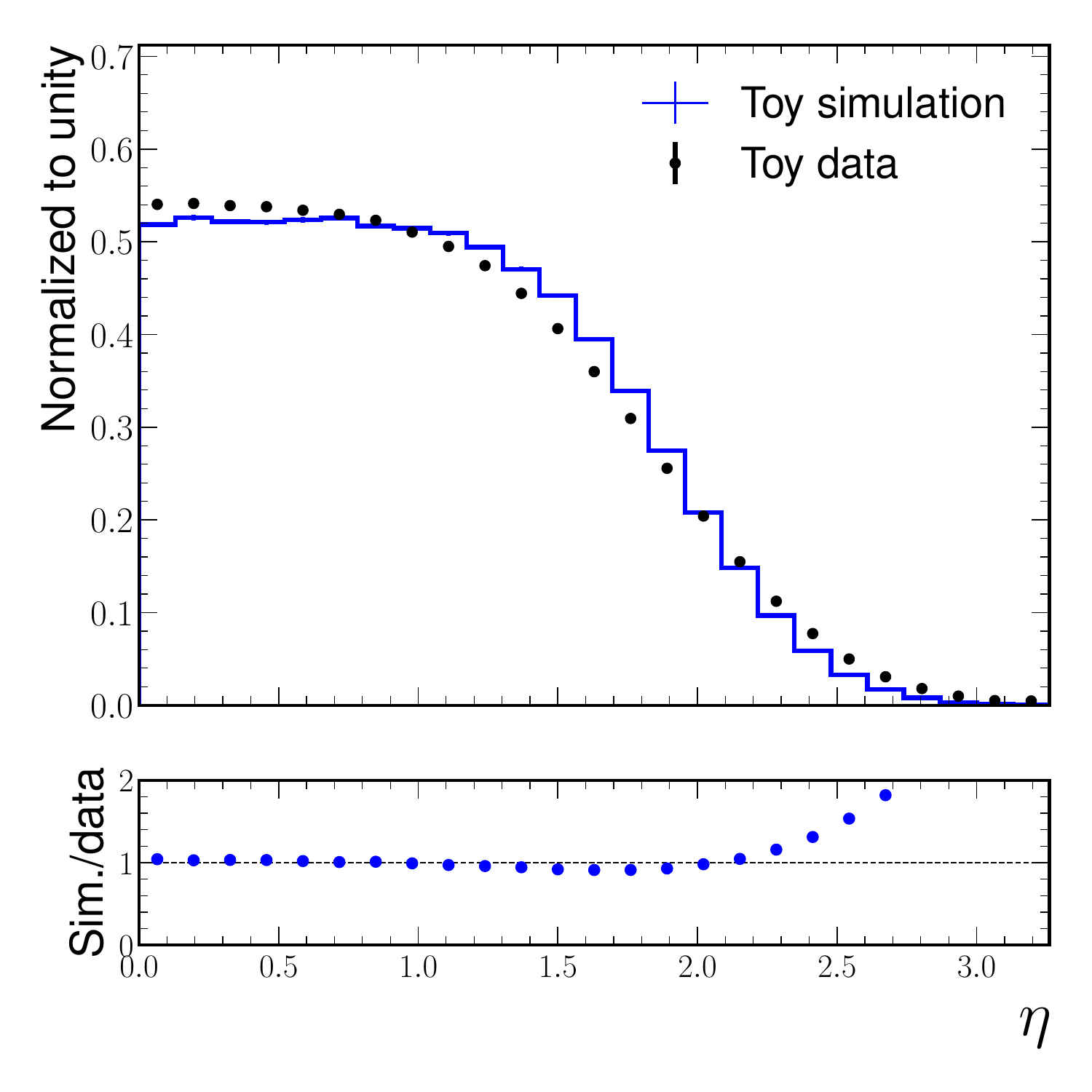}
    \end{subfigure}
    \\
    \begin{subfigure}{0.33\textwidth}
        \includegraphics[width=\linewidth]{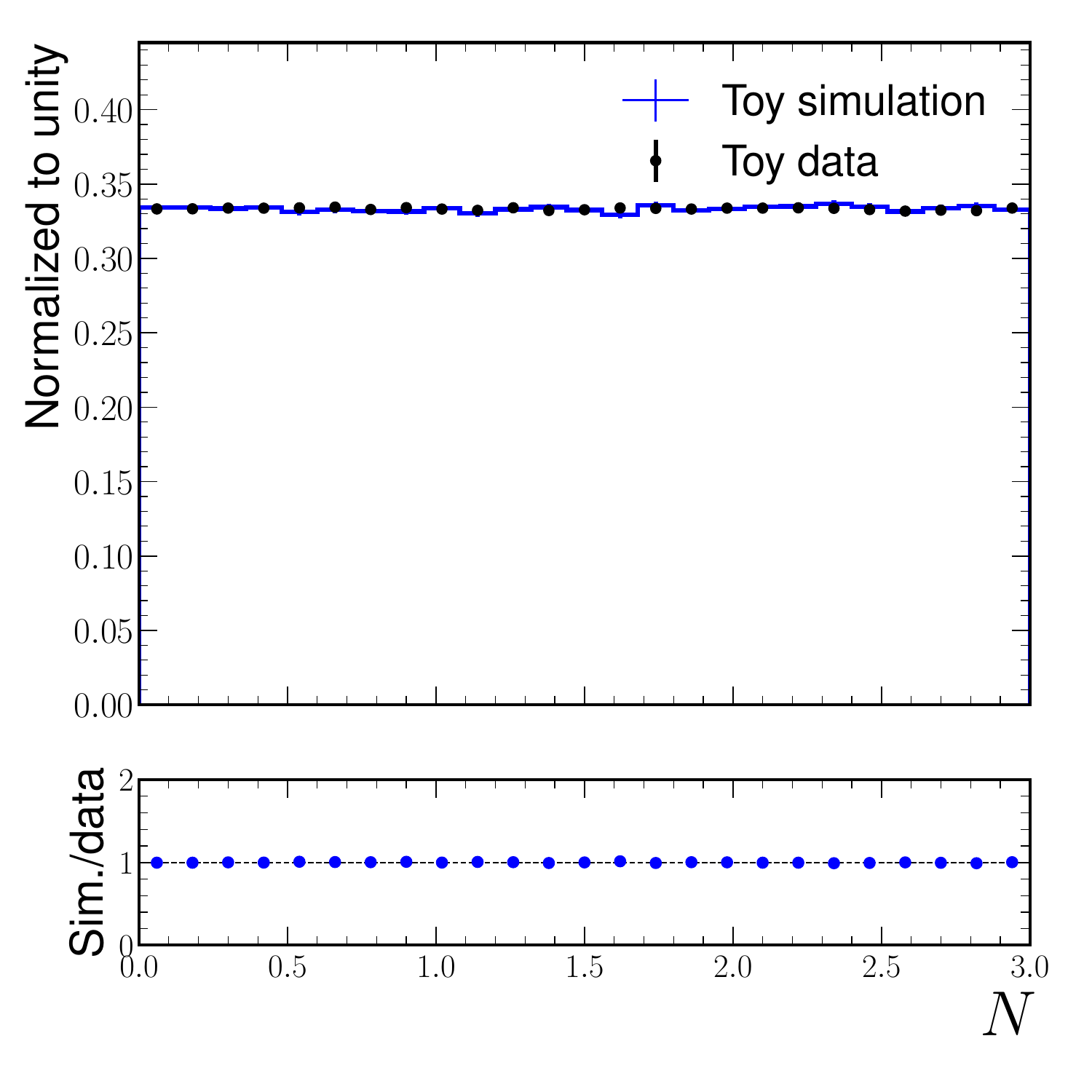}
    \end{subfigure}
    \\
    \begin{subfigure}{0.33\textwidth}
        \includegraphics[width=\linewidth]{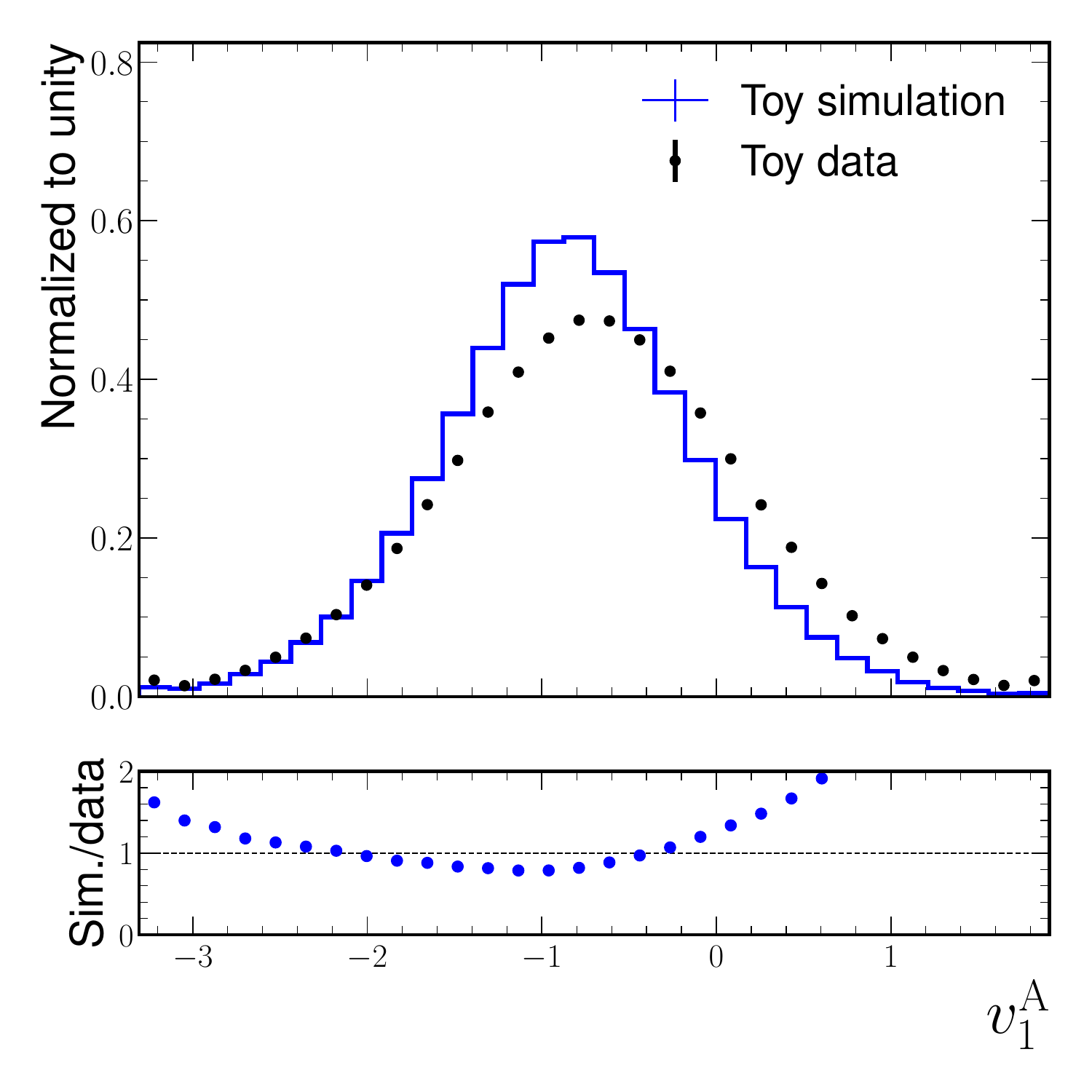}
    \end{subfigure}
    \hspace{1cm}
    \begin{subfigure}{0.33\textwidth}
        \includegraphics[width=\linewidth]{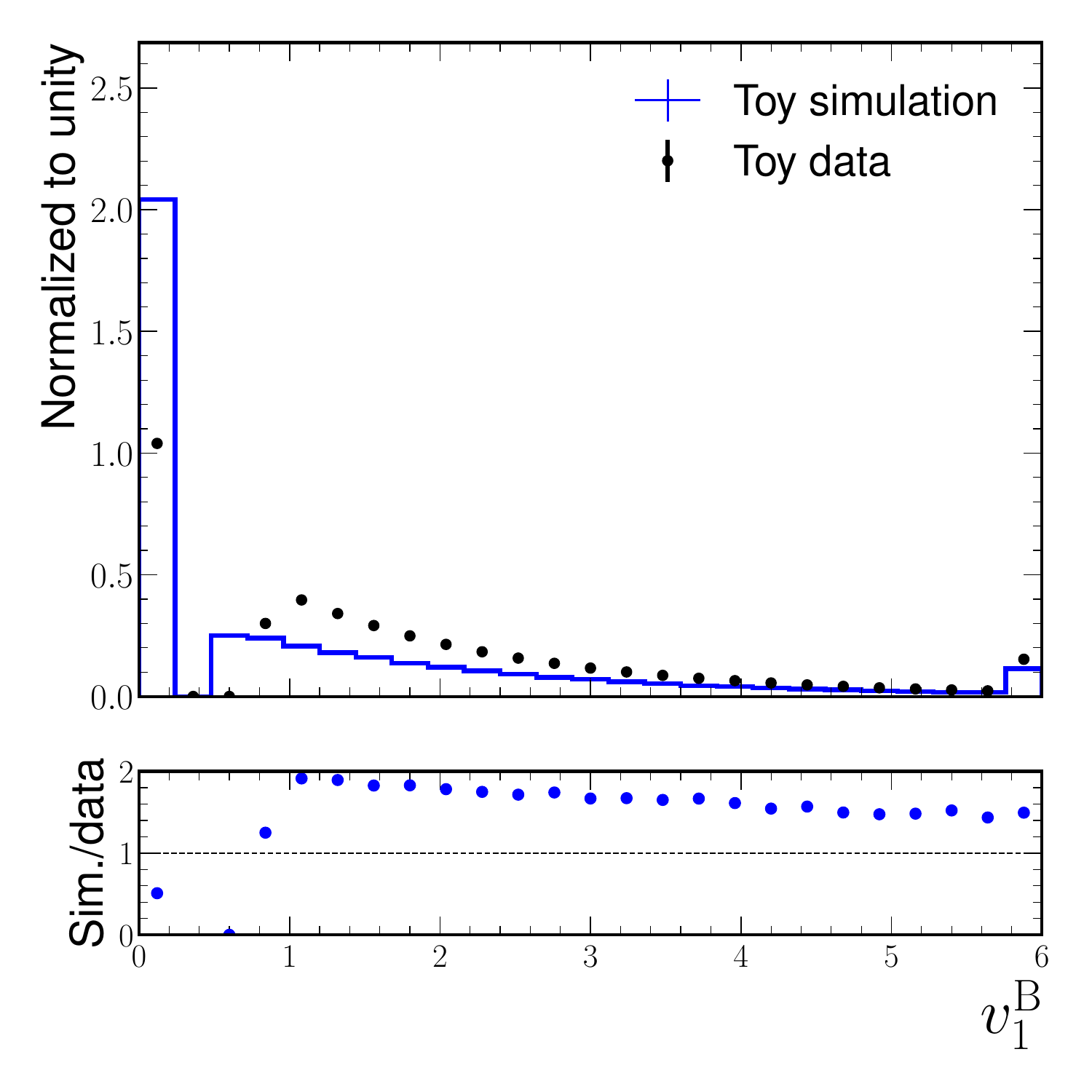}
    \end{subfigure}
    \\
    \begin{subfigure}{0.33\textwidth}
        \includegraphics[width=\linewidth]{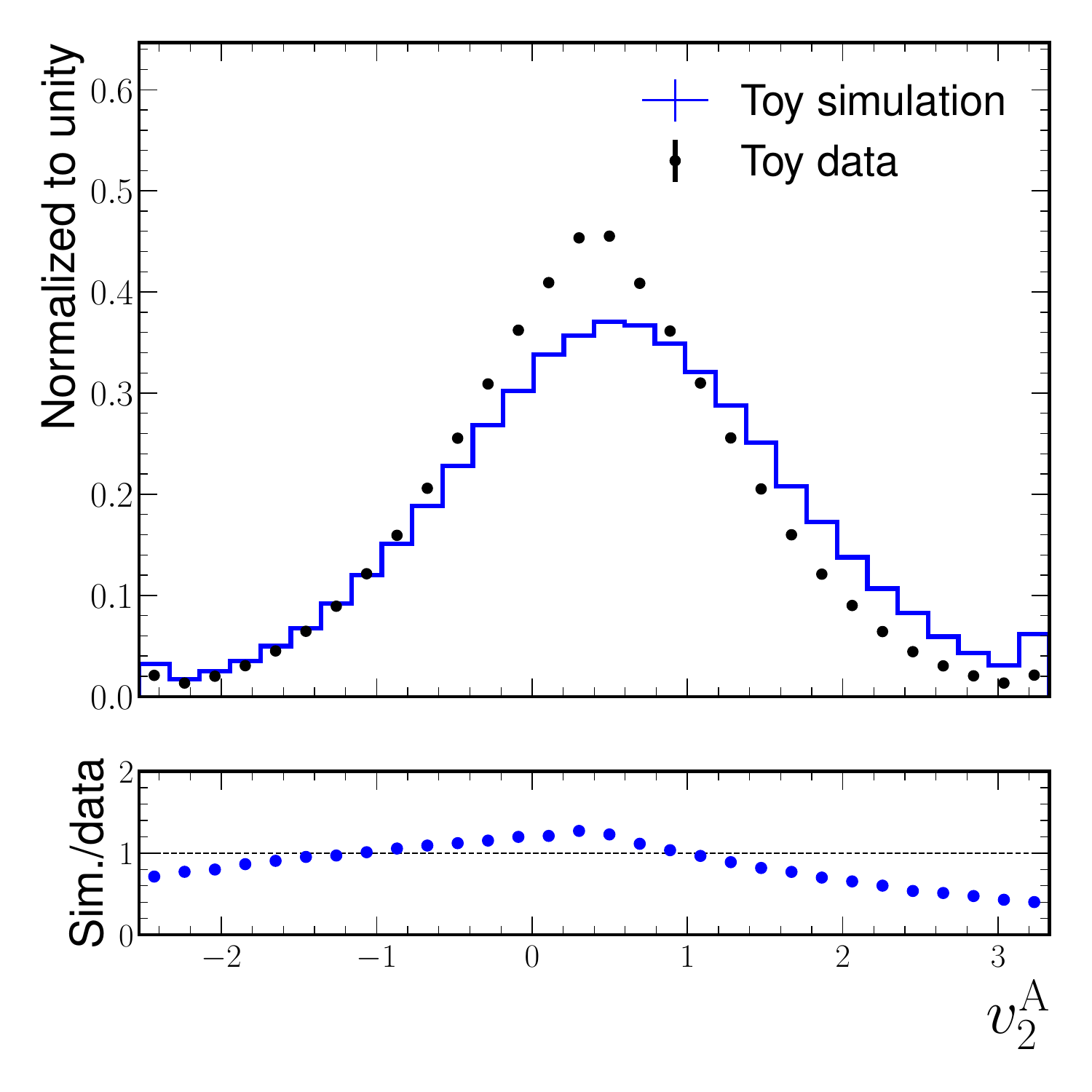}
    \end{subfigure}
    \hspace{1cm}
    \begin{subfigure}{0.33\textwidth}
        \includegraphics[width=\linewidth]{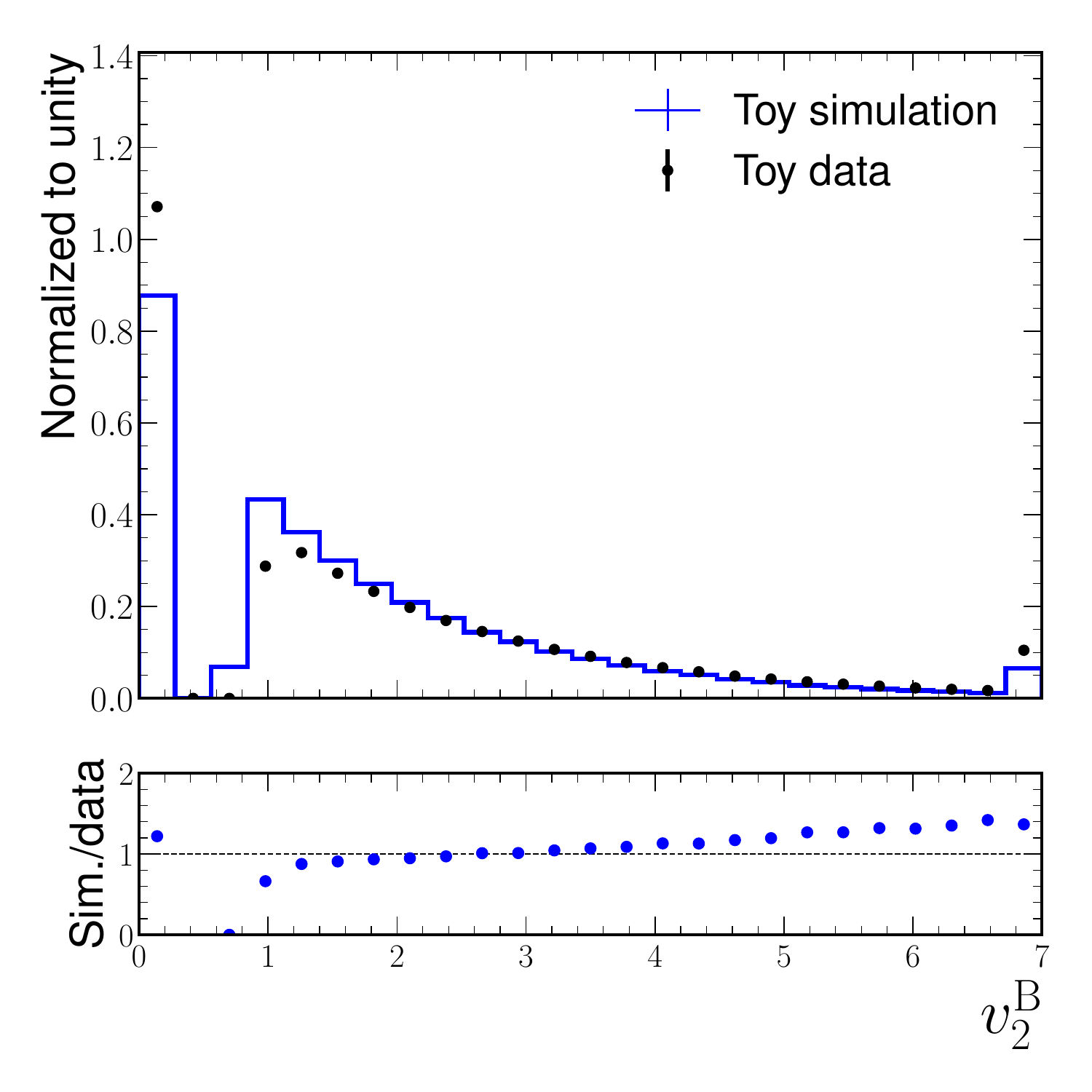}
    \end{subfigure}
    \caption{Marginal distributions for the seven variables in the data and simulation datasets. The three ancillary variables are shown in the upper figures and the four informative features in the lower figures.
    The ancillary variables are defined in a way that \pt\ is unitless, \myEta\ takes only positive values and \noise\ is defined in the interval [0,3].
    }
    \label{fig:datasets}
\end{figure}

As a final step, the seven variables are endowed with non-trivial correlations using the \texttt{mcerp} package~\cite{mcerp,mcerp_theory}.
While the marginal distributions before and after the artificial correlation are very similar, we are able to impose non-trivial correlations between all features, which increases the complexity of the dataset.
The resulting correlation matrices differ significantly between data and simulation.
Figure~\ref{fig:datasets} shows the marginal distributions of the seven variables for data and simulation.
Two-dimensional visualisations with $68\%$ contours are given in Appendix~\ref{app:cornerplot}.

We generate two statistically independent datasets for toy data and toy simulation, respectively, with ten million events each.
More technical details for the generation of the dataset are given in Appendix~\ref{app:gen_formulas}. The code for generating the dataset is publicly available\footnote{\url{https://github.com/JaLuka98/OneFlowPublic}}.

\FloatBarrier

\section{Training and results on the physics-inspired dataset}
\label{sec:performance}

\subsection{Preprocessing and training}
\label{subsec:arch_prepro_train}

\begin{figure}[h]
    \centering
    \includegraphics[clip, trim=13.5cm 13cm 16.5cm 1cm,width=0.65\textwidth]{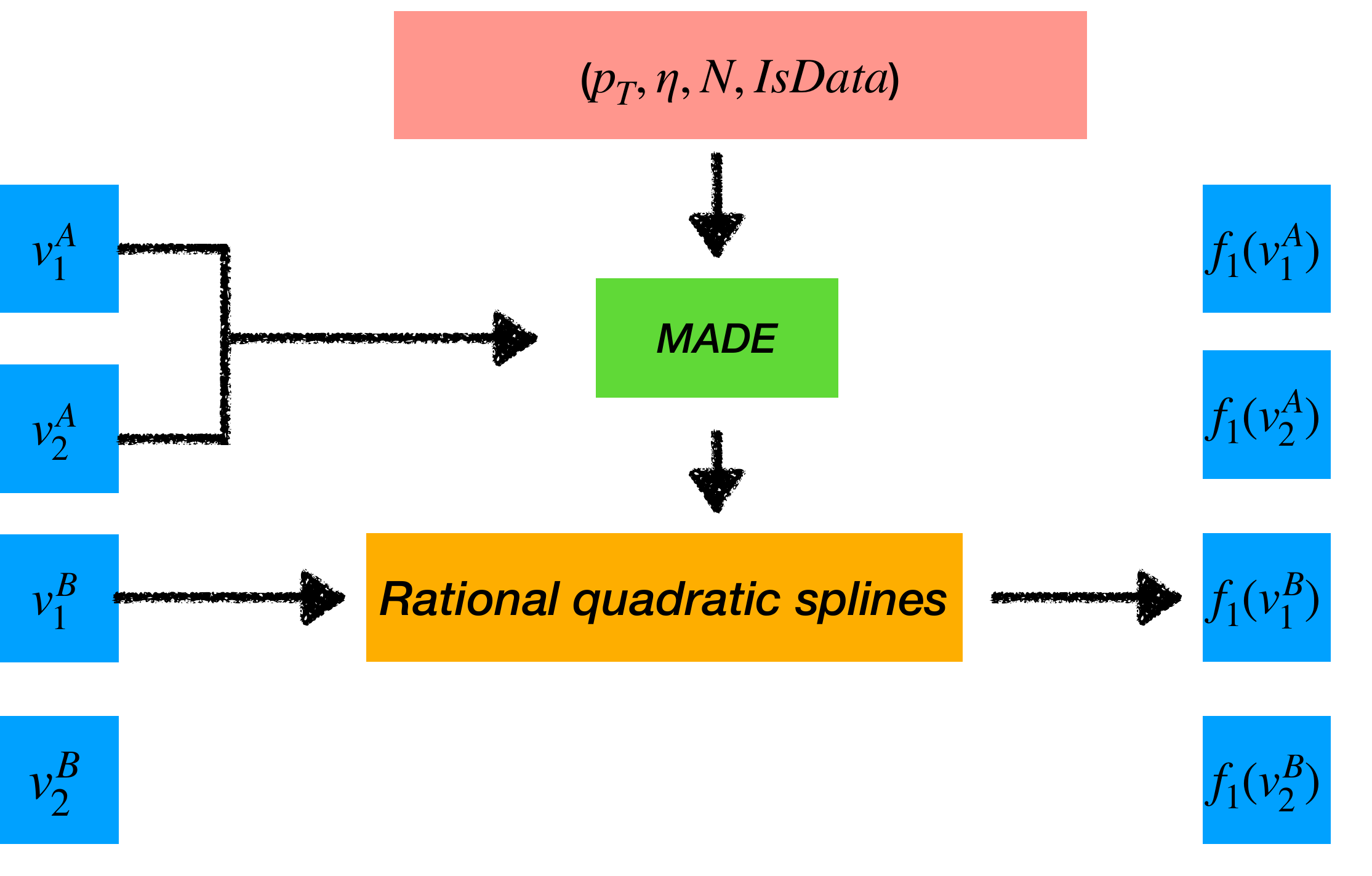}
    \caption{Illustration of the forward pass of the normalising flow for the example of informative feature~\vOneB.
    The four informative features are transformed by the autoregressive structure one at a time.
    The ancillary variables and the IsData boolean are conditional inputs to the Masked Autoencoder for Distribution Estimation (MADE) neural network, which generates the parameters for the rational quadratic splines that transforms the variables.
    }
    \label{fig:sketch}
\end{figure}

Figure~\ref{fig:sketch} illustrates the forward pass of the normalising flow for the example of the informative feature~\vOneB.
In the forward pass, all four informative features are transformed to the four-dimensional base distribution via autoregressive rational quadratic splines, as described in Section~\ref{sec:corrections}.
The MADE blocks are conditioned on the IsData boolean and the three ancillary variables \pt, \myEta\ and \noise.
\textcolor{black}{Thus, the learning objective is the conditional probability $p(\vOneA, \vOneB, \vTwoA, \vTwoB \mid \pt, \myEta, \noise, \mathrm{isData})$.}
\textcolor{black}{This setup corresponds to the typical application of this correction method in high-energy physics, where the model is trained in a control region and applied in a signal region with different distributions of the ancillary features.}

The preprocessing of the input variables includes a reweighting step for the ancillary variables and a smoothing step for discontinuous informative features.
Additionally, all variables are studentised, i.e.~we use $(v_i-\bar{v})/s_{v}$, with $\bar{v}$ and $\sigma_v$ the sample mean and the sample standard deviation of the given sample of variable $v$, respectively.
The reweighting and smoothing procedures are introduced below.

The reweighting step ensures that the distributions of the ancillary variables match well between data and simulation.
The ancillary variables are used as conditions for the normalising flow so that the correction is determined as a function of these variables.
The idea is that the corrections can be used in samples that show a different distribution in the ancillary variables than the distribution of the training sample.
Thus, the reweighting step avoids effects from differences between data and simulation in the ancillary variables.
The reweighting is performed simultaneously in the three ancillary variables with $16$~bins for \pt, $16$~bins \myEta\ for and $10$~bins for \noise.
The binning is chosen such that each bin contains approximately the same number of events.

Discontinuous informative features are subject to a smoothing step, making them continuous.
The reason is that normalising flows use differentiable transformations to map the original distribution to a differentiable base distribution.
As this can only be approximate for discontinuous input distributions, the overall performance of normalising flows for the morphing application may suffer in such cases~\textcolor{black}{\cite{PhysRevD.107.113003_CALOflow}}.
In our dataset, this applies to the variables \vOneB\ and \vTwoB.

\begin{figure}[t]
    \centering
    \begin{subfigure}{1.0\textwidth}
        \includegraphics[width=\linewidth]{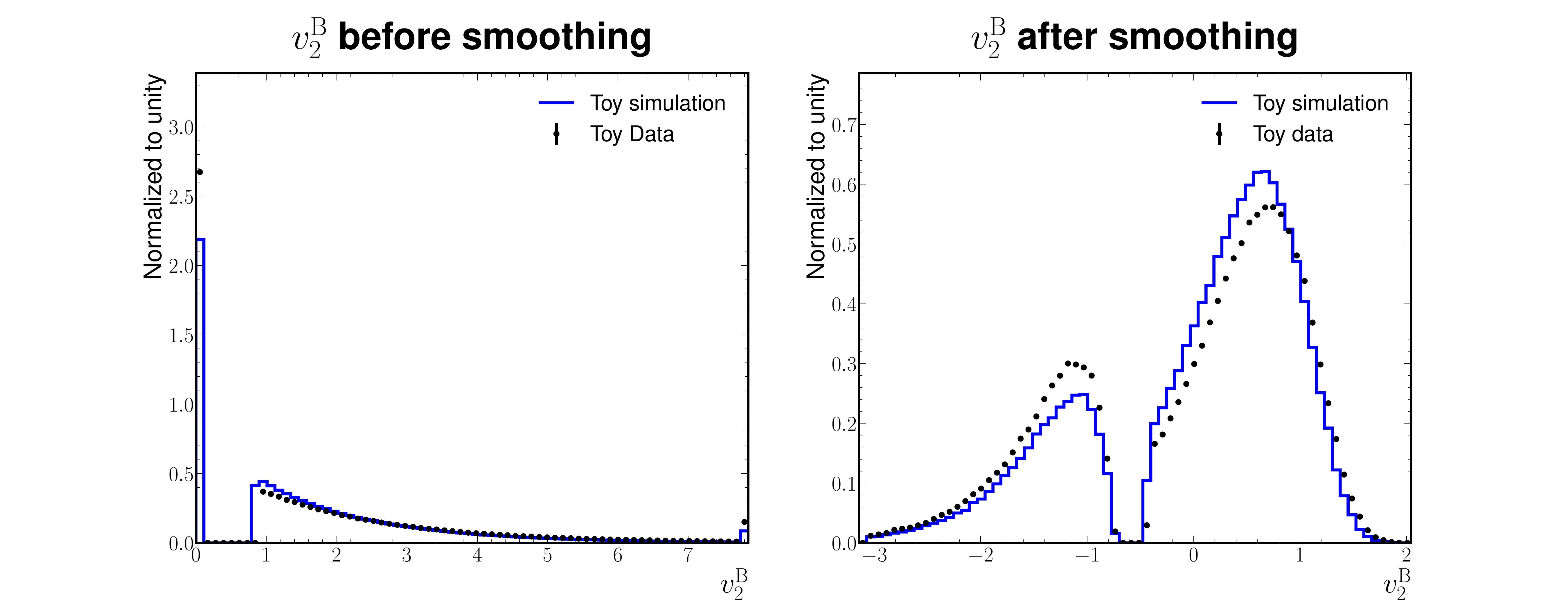}
    \end{subfigure}
    \caption{Distribution of the informative feature \vTwoB\ before~(left) and after the smoothing and the logarithmic transformation~(right) for nominal simulation and data, normalised to unit area. 
    The first bin in the distribution on the right includes the underflow. The last bin in both distributions includes the overflow.}
    \label{fig:iso_transform}
\end{figure}

The smoothing step aims to fill gaps between continuous parts of distributions.
In our dataset, the discontinuity consists of a peak at zero that is followed by a gap and a smoothly falling tail.
Figure~\ref{fig:iso_transform}~(left) shows the feature \vTwoB\ as an example.
We substitute any zero value with a random value according to a triangular distribution with a fixed slope and then move the tail close to the end point of the triangle.
We chose a triangular distribution for simplicity and because it fulfils the requirement to not overlap with the continuous tail so that the inversion of the random sampling is unambiguous.
In the case of our dataset, we include an additional logarithmic transformation that spreads the values within the previous gap closer to the tail.
The result of the smoothing step for \vTwoB\ is shown in Figure~\ref{fig:iso_transform}~(right).
 
As a final preprocessing step, all four informative features and the three ancillary variables are studentised.
This also applies to the smoothed and transformed \viB\ variables.

As for the benchmark datasets in Section~\ref{sec:corrections}, the generated datasets are again separated into training, validation and test datasets with a split of $60\%$, $10\%$ and $30\%$.
The whole dataset is composed of 2.5~million data samples and 2.5~million simulation samples.
We do not use the entire generated dataset, as this is not necessary for the successful training of the model.

A normalising flow as introduced in Section~\ref{sec:corrections}, and illustrated in Figure~\ref{fig:sketch}, is used for to correct the simulation in the physics-inspired dataset using six monotonic rational-quadratic transformations.
We did not perform a detailed hyperparameter optimisation, as we observed stable and satisfactory results without extensive tuning.
As a cross-check, we trained a model with double the number of transformations and used three layers in the auxiliary networks, but we did not observe any significant differences in the performance of the corrections.
Early stopping is used to end the training process once the loss on the validation dataset fails to improve for $15$~consecutive epochs.

\subsection{Evaluation of the corrections}
\label{subsec:evaluation_corrections}

We evaluate the quality of the corrections by checking the agreement between data and corrected simulations in (a)~the marginal distributions of the four informative features, (b)~in the Pearson correlation coefficients between all seven variables in the dataset, and (c)~in the output distribution of a boosted decision tree (BDT) that is trained to distinguish between data and corrected simulations.

\begin{figure}[p]
    \centering
        \includegraphics[width=0.49\textwidth]{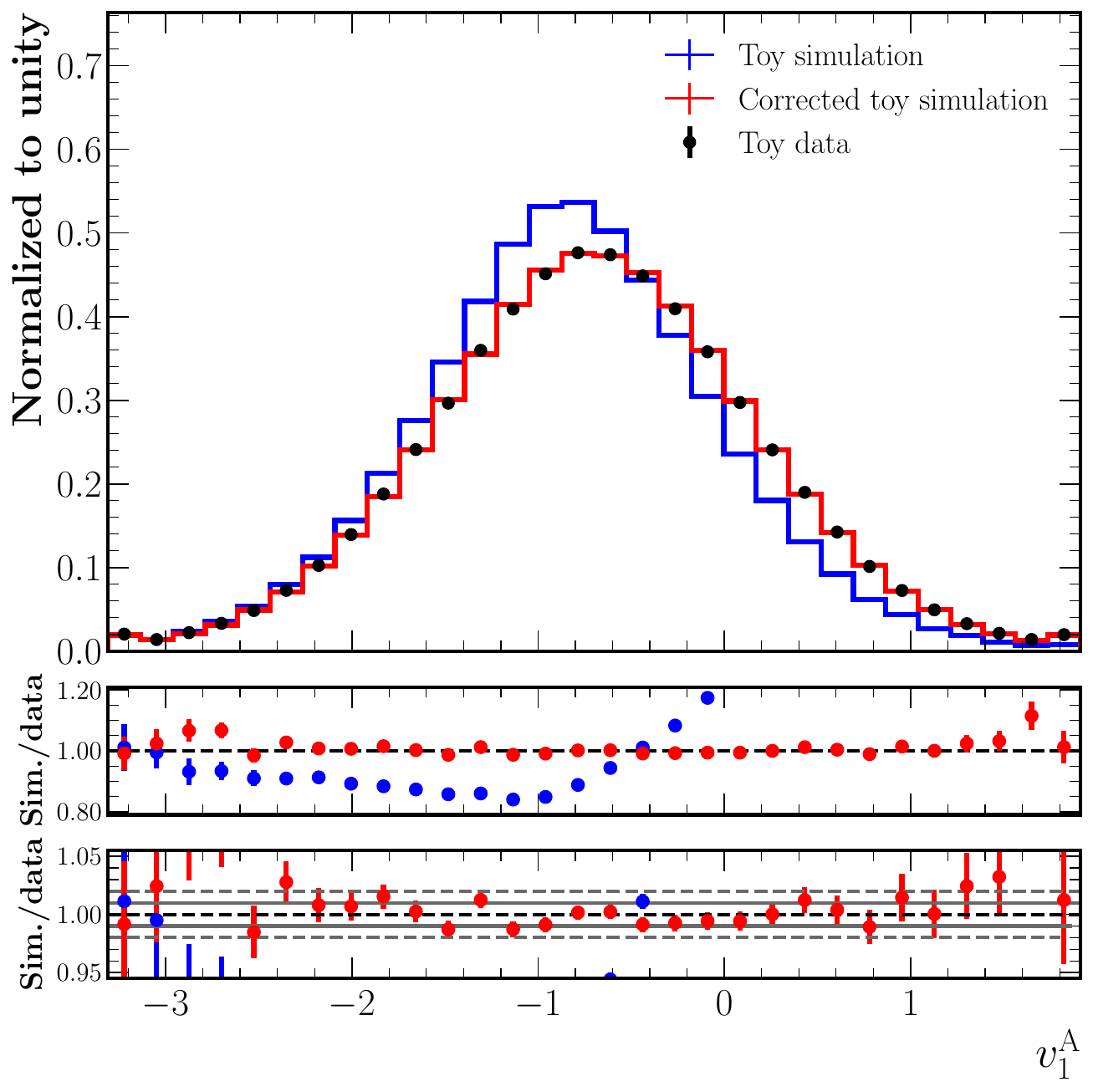}
        \includegraphics[width=0.49\textwidth]{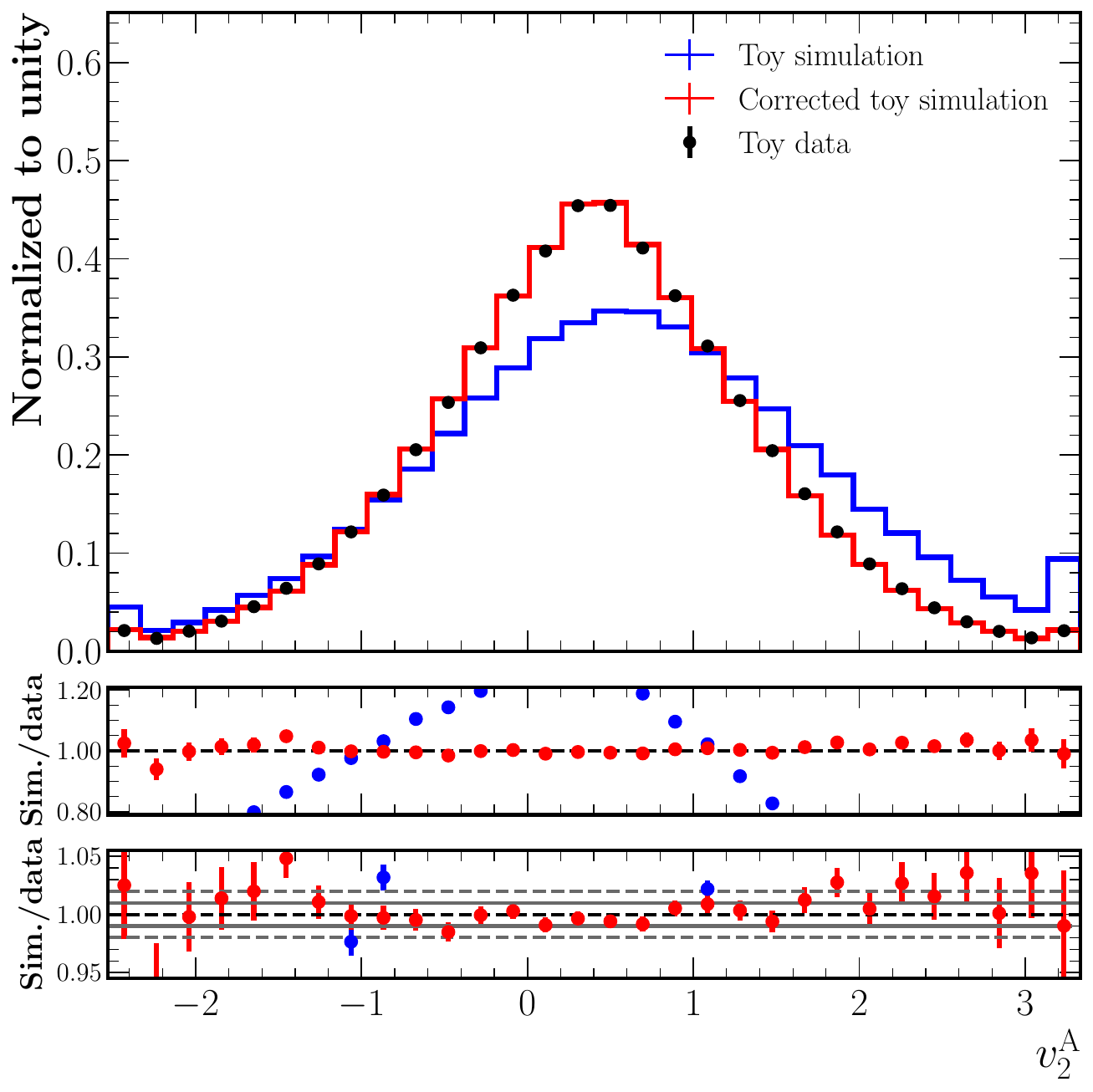}
        \includegraphics[width=0.49\textwidth]{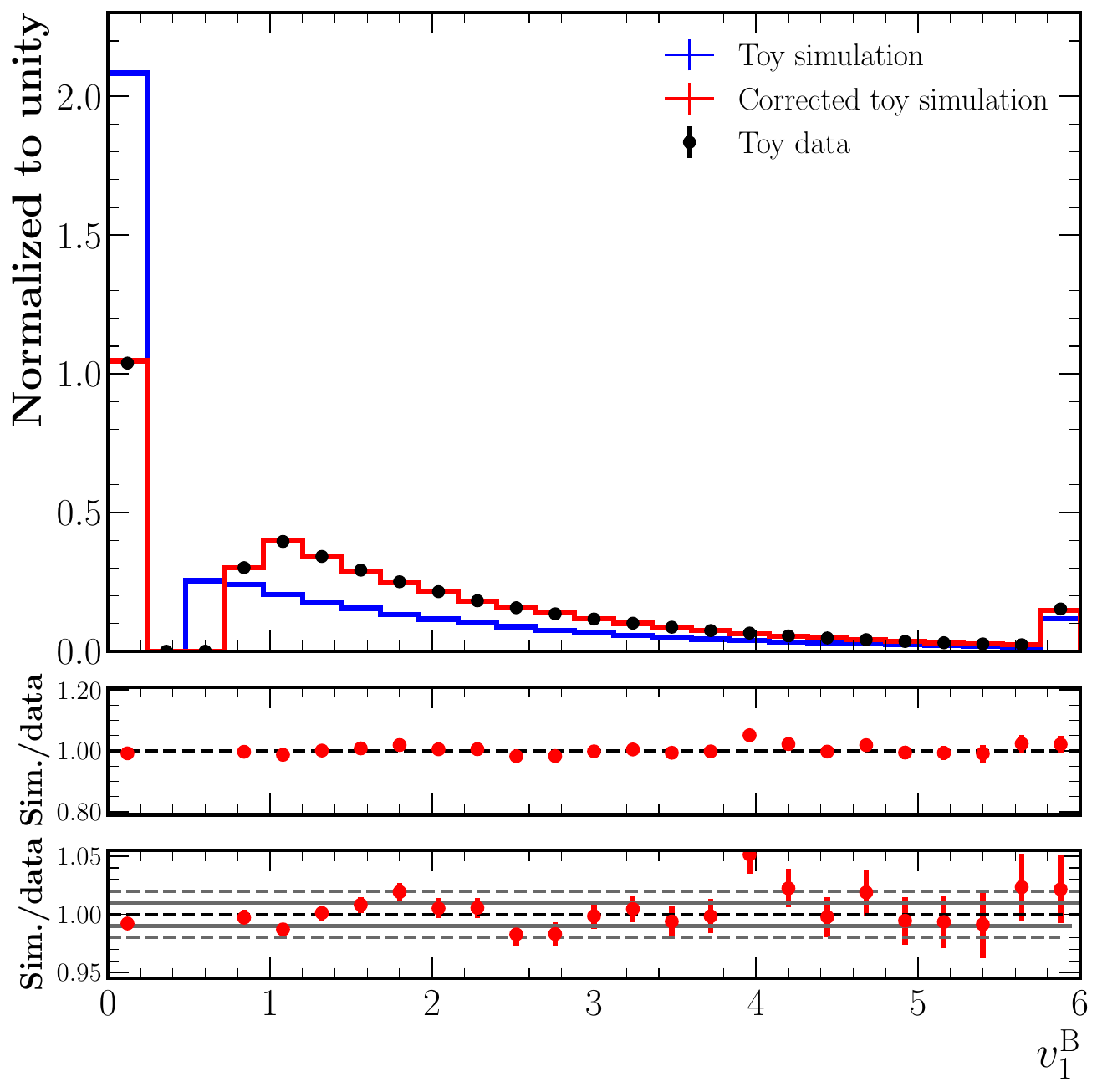}
        \includegraphics[width=0.49\textwidth]{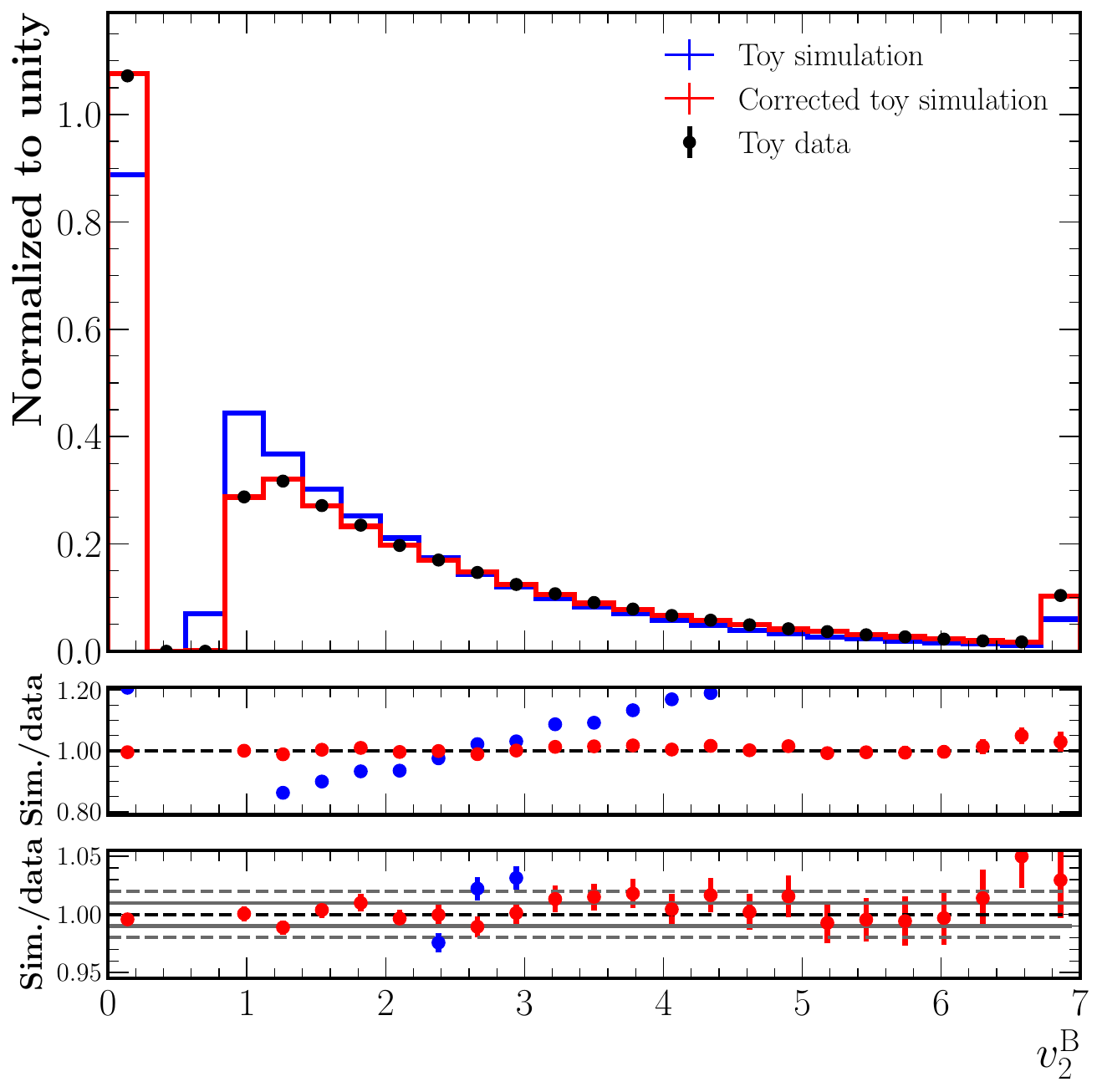}
    \caption{Nominal simulation (blue), corrected simulation (red) and data (black) marginal distributions (normalised to unit area) for the four informative features.
    The two panels below each figure show the ratio of nominal/corrected simulation and the data with a range of $1.0\pm 0.2$ and with a closer zoom of $1.00\pm 0.05$.
    Markers that are not shown in the ratio plots are out of the $y$-axis range.
    The first bin in the upper distributions includes the underflow. The last bin in all distributions includes the overflow.
    }
    \label{fig:flow_marginal_dists}
\end{figure}

The marginal distributions of the four informative features for uncorrected and corrected simulations and the data are shown in Figure~\ref{fig:flow_marginal_dists}.
The distributions are normalised to unit area.
The agreement between simulation and data is strongly improved by the normalising-flow corrections.
We observe that the agreement after the correction is at the level of $1$-$2\%$ in the bulk of the distributions for all four informative features.
In the tails of the distributions, where the uncertainties due to the limited size of the test dataset (calculated from the variance of the sum of squared event weights) are of the order of a few per cent, the corrected simulation still agrees with the data within these uncertainties.
The very good agreement in the discontinuous distributions \vTwoA\ and \vTwoB\ illustrates that the smoothing step in the preprocessing of these variables was successful.
This is especially true for \vOneB, where a substantial migration is required from the peak at zero to the tail of the distribution for the simulation and the data to match.

\begin{figure}[p]
    \centering
    \begin{subfigure}{0.49\textwidth}
        \includegraphics[width=\linewidth]{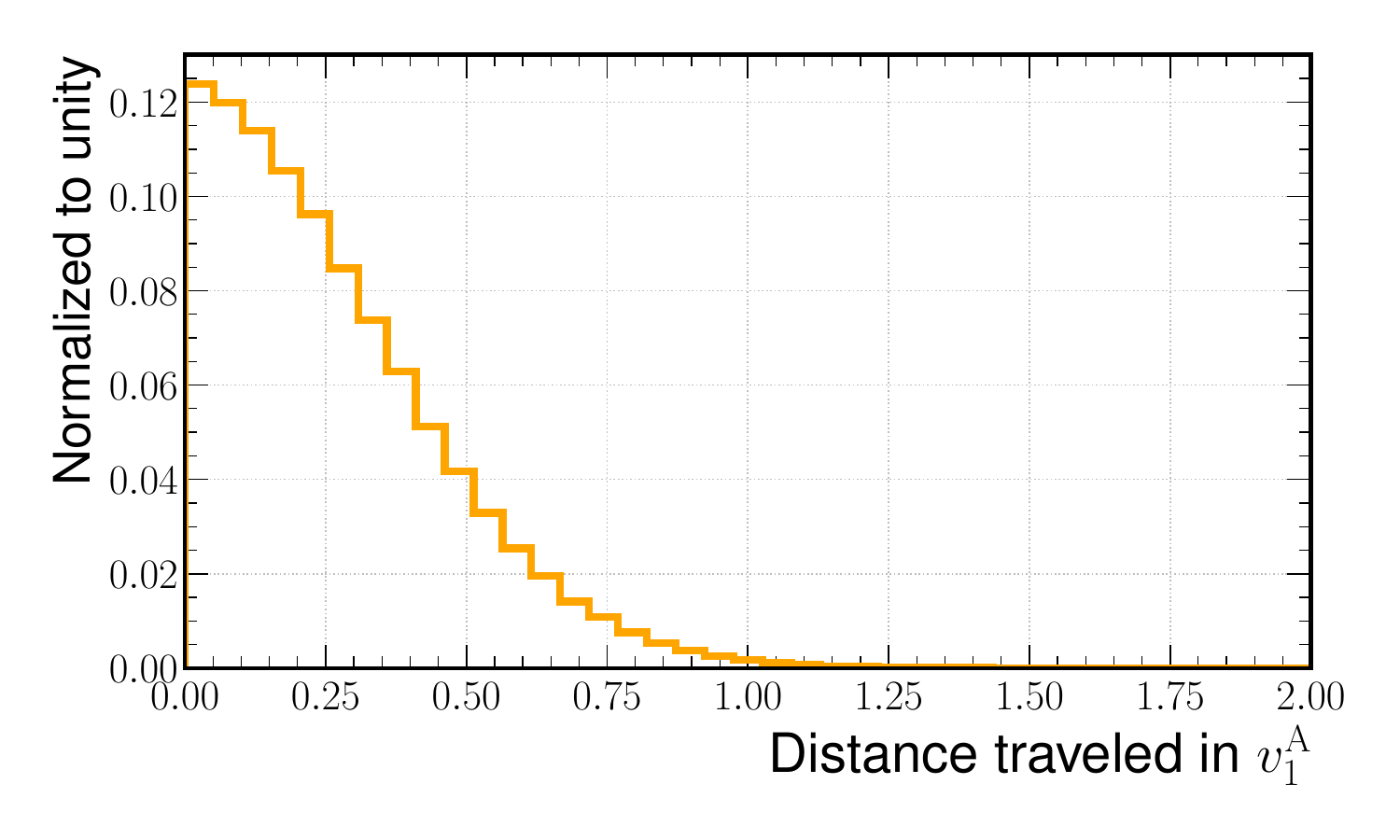}
    \end{subfigure}
    \hfill
    \begin{subfigure}{0.49\textwidth}
        \includegraphics[width=\linewidth]{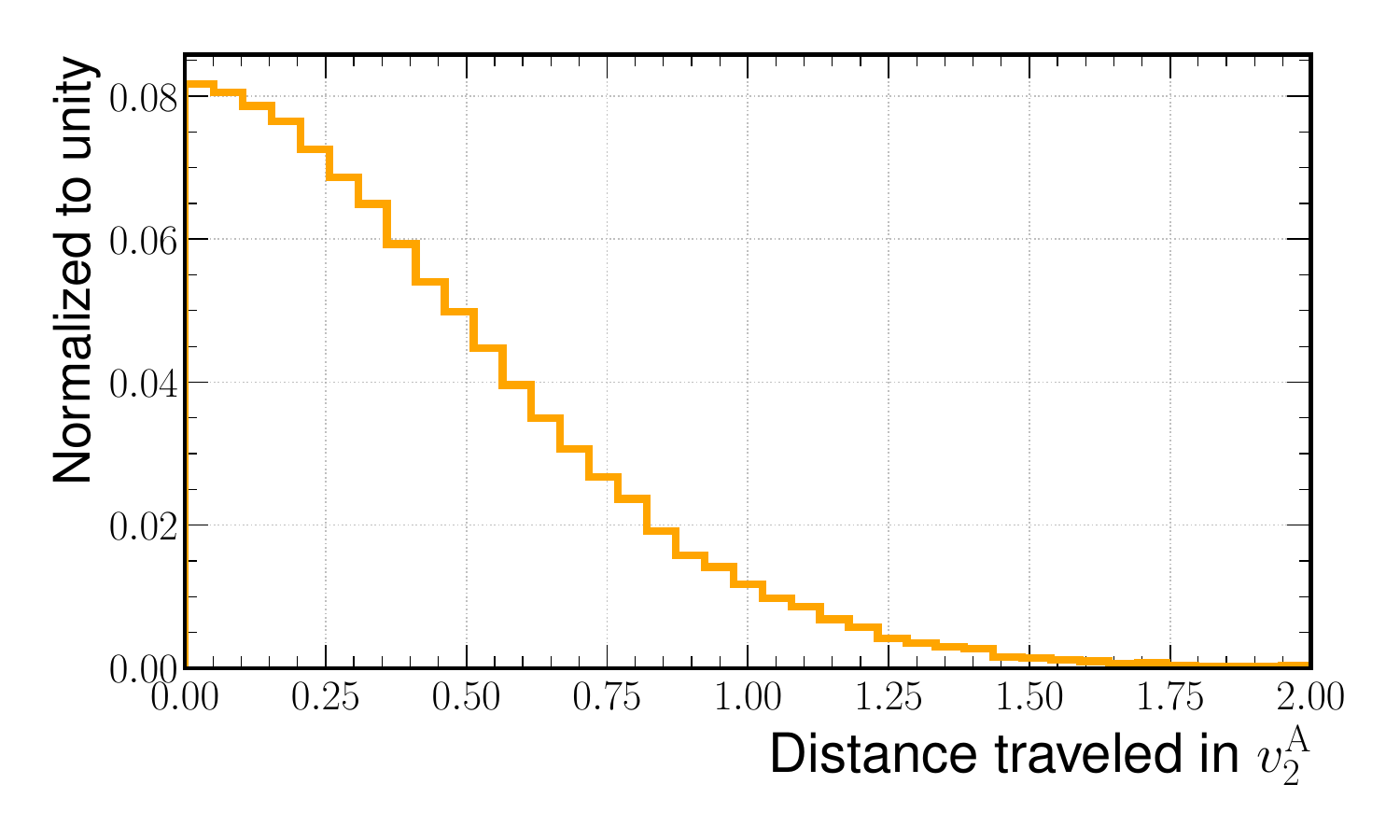}
    \end{subfigure}
    \hfill
        \begin{subfigure}{0.49\textwidth}
        \includegraphics[width=\linewidth]{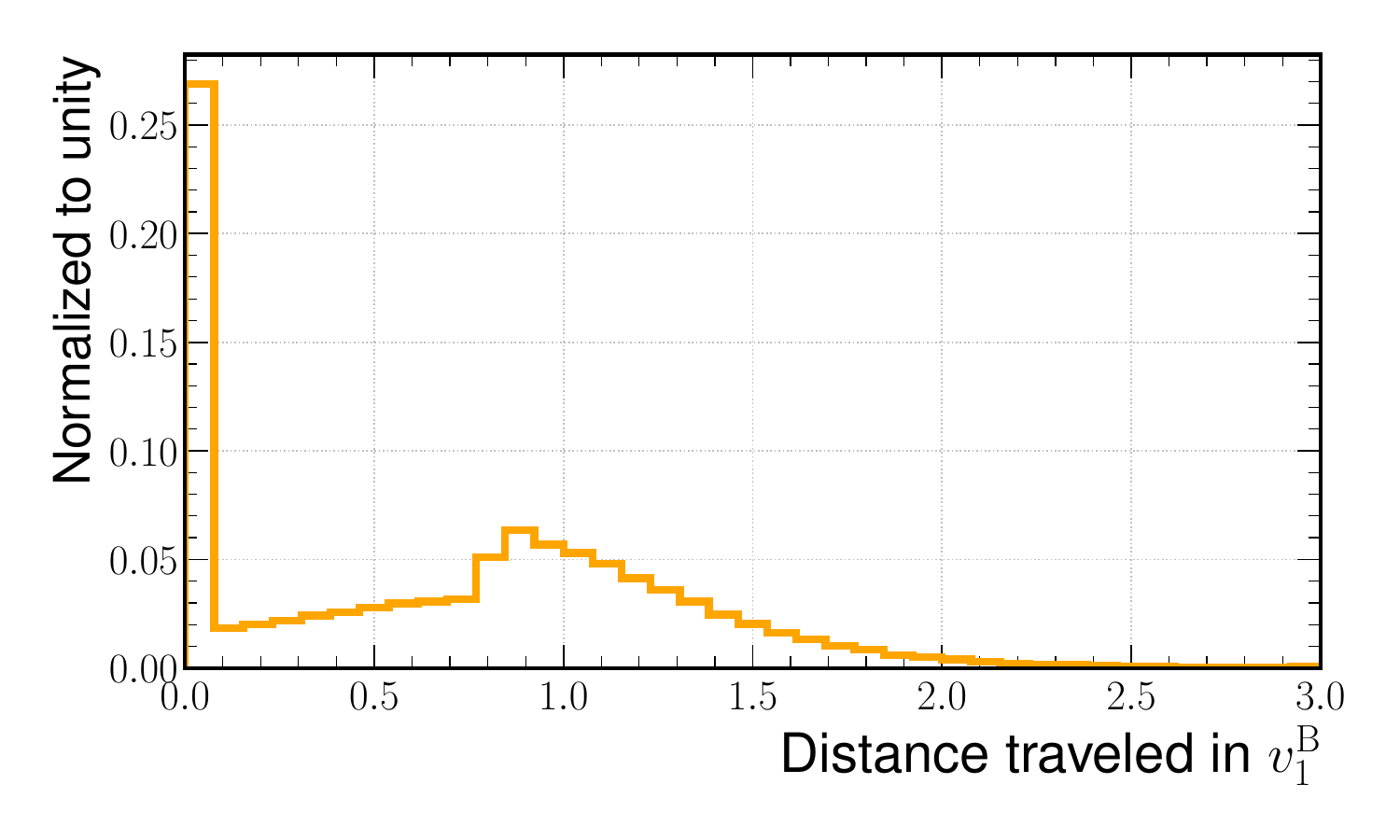}
    \end{subfigure}
    \hfill
    \begin{subfigure}{0.49\textwidth}
        \includegraphics[width=\linewidth]{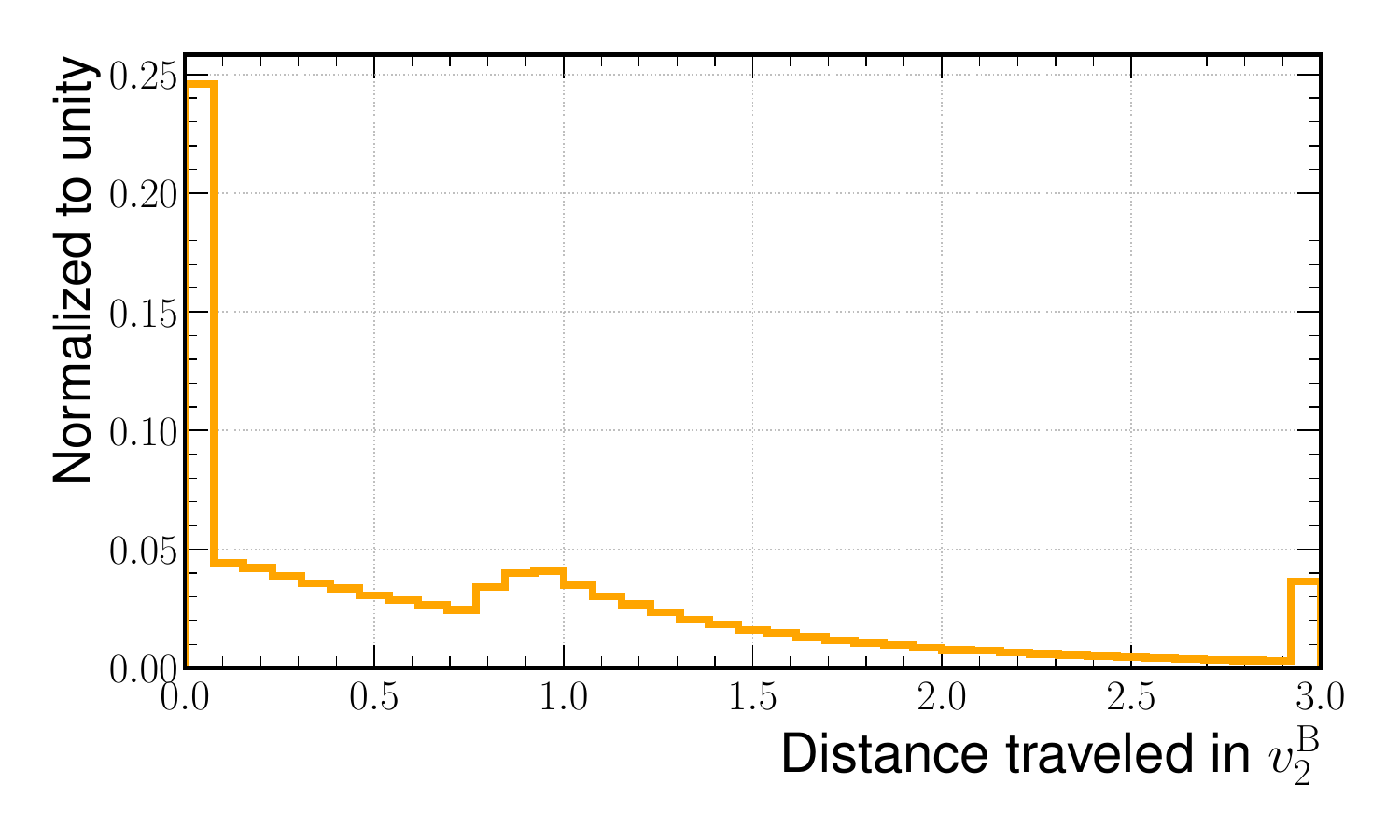}
    \end{subfigure}

    \hfill
    \caption{Marginal distributions of the corrections applied to simulation for the four informative features (``distance traveled'').
    }
    \vspace{1cm}
    \label{fig:distances}
        \includegraphics[width=0.45\textwidth]{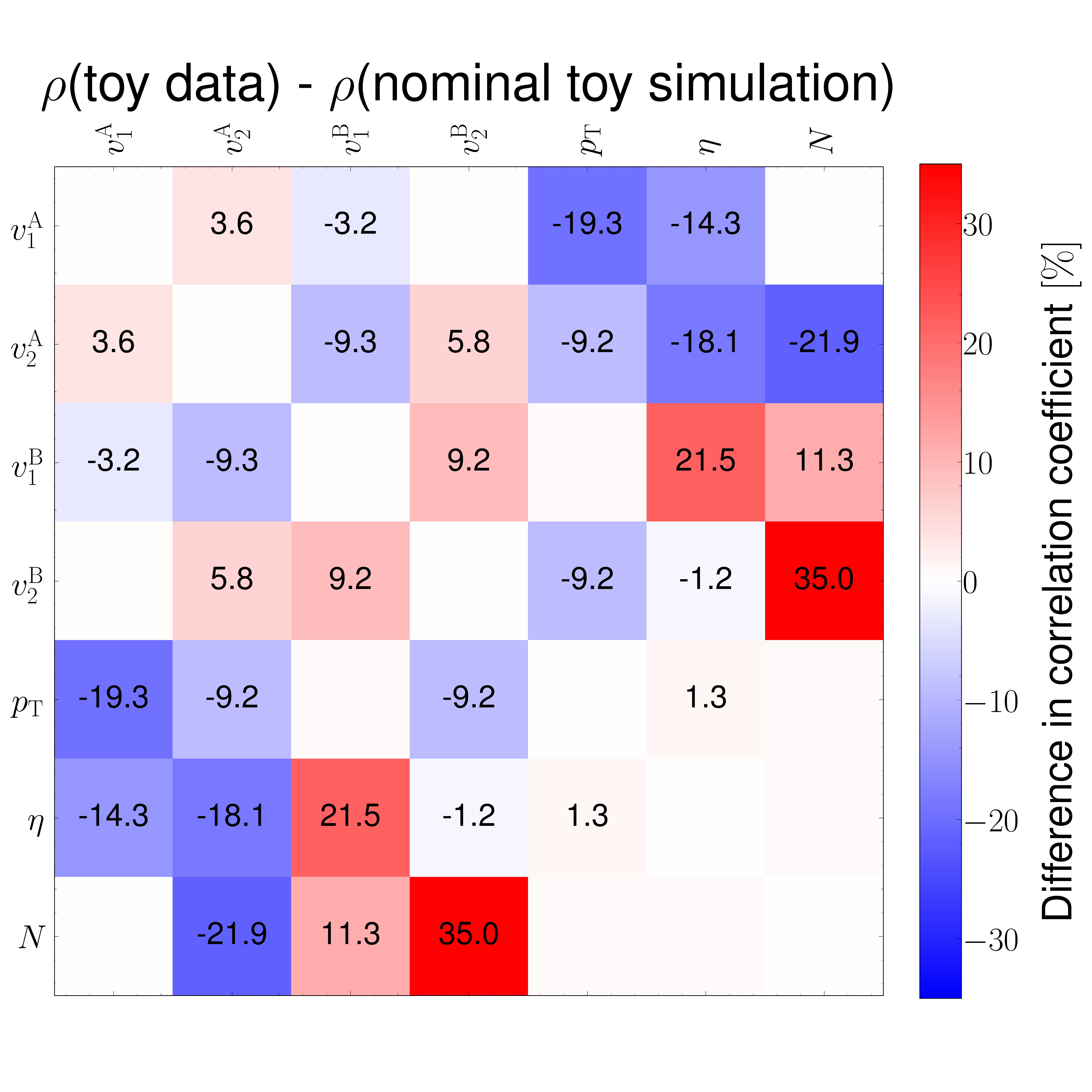}
        \label{fig:corr_matrix_data_uncorrected}
    \hspace{1cm}
        \includegraphics[width=0.45\textwidth]{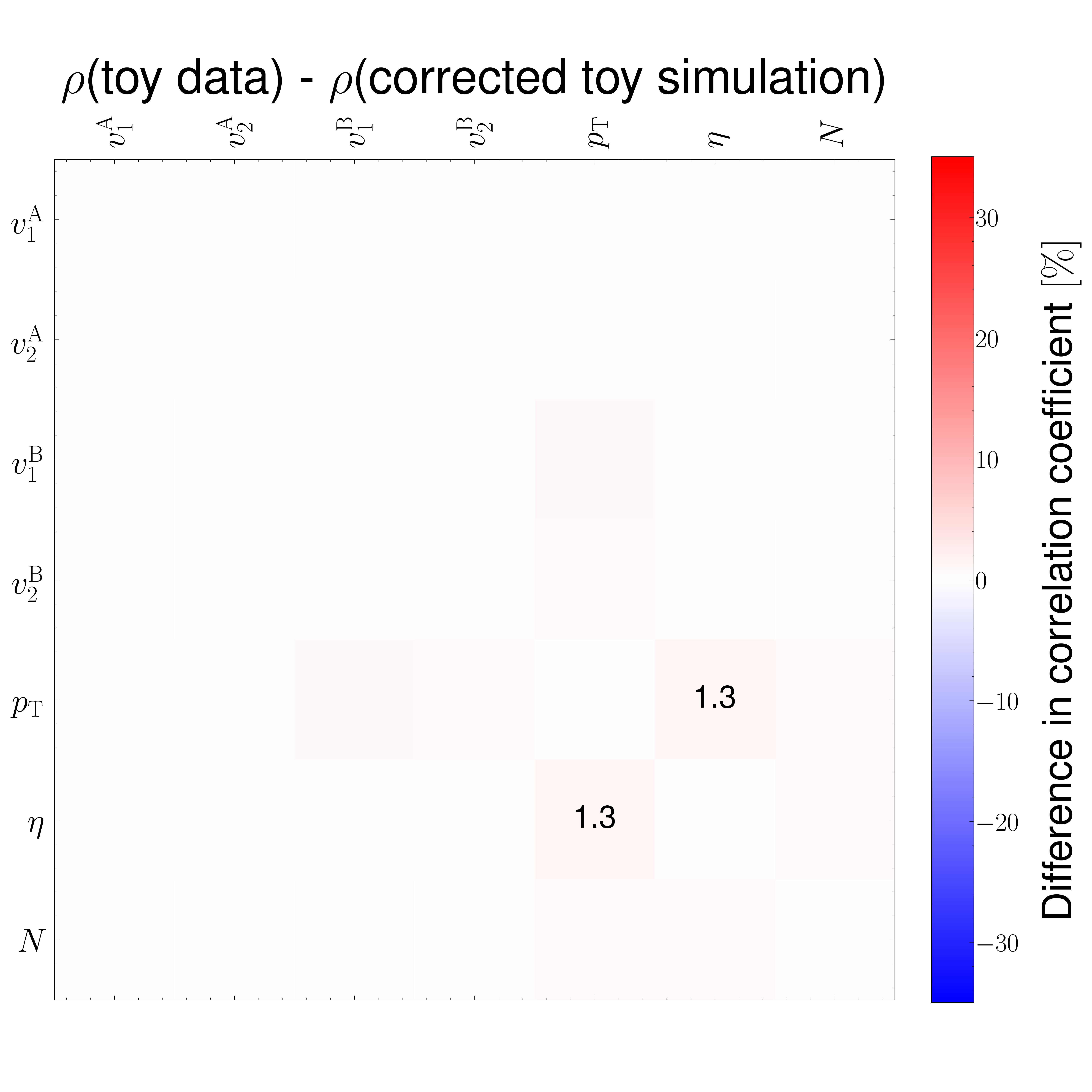}
        \label{fig:corr_matrix_data_corrected}
    \caption{Differences of the Pearson correlation coefficients ($\rho$) between nominal toy simulation and toy data (left) and between corrected simulation and data (right). The $\rho$ values are given in per cent. Values of $\rho$ with $|\rho|<1\%$ are not shown as numbers but are only encoded by the bin colour.
    }
    \label{fig:corr_matrices}
\end{figure}

Figure~\ref{fig:distances} shows the marginal distributions of the corrections applied to simulation for the four informative features.
The continuous variables, \vOneA\ and \vTwoA, show a smoothly falling distribution with the maximum at zero.
The discontinuous variables, \vOneB\ and \vTwoB, show a sharp peak at zero and then a second maximum.
While the peak at zero comes from samples that were not moved in \vTwoA\ or \vTwoB, respectively, the second maximum originates from migrations from the original peaks to the tails of the distributions and vice versa.

\begin{figure}[p]
    \centering
        \centering 
        \includegraphics[width=0.49\textwidth]{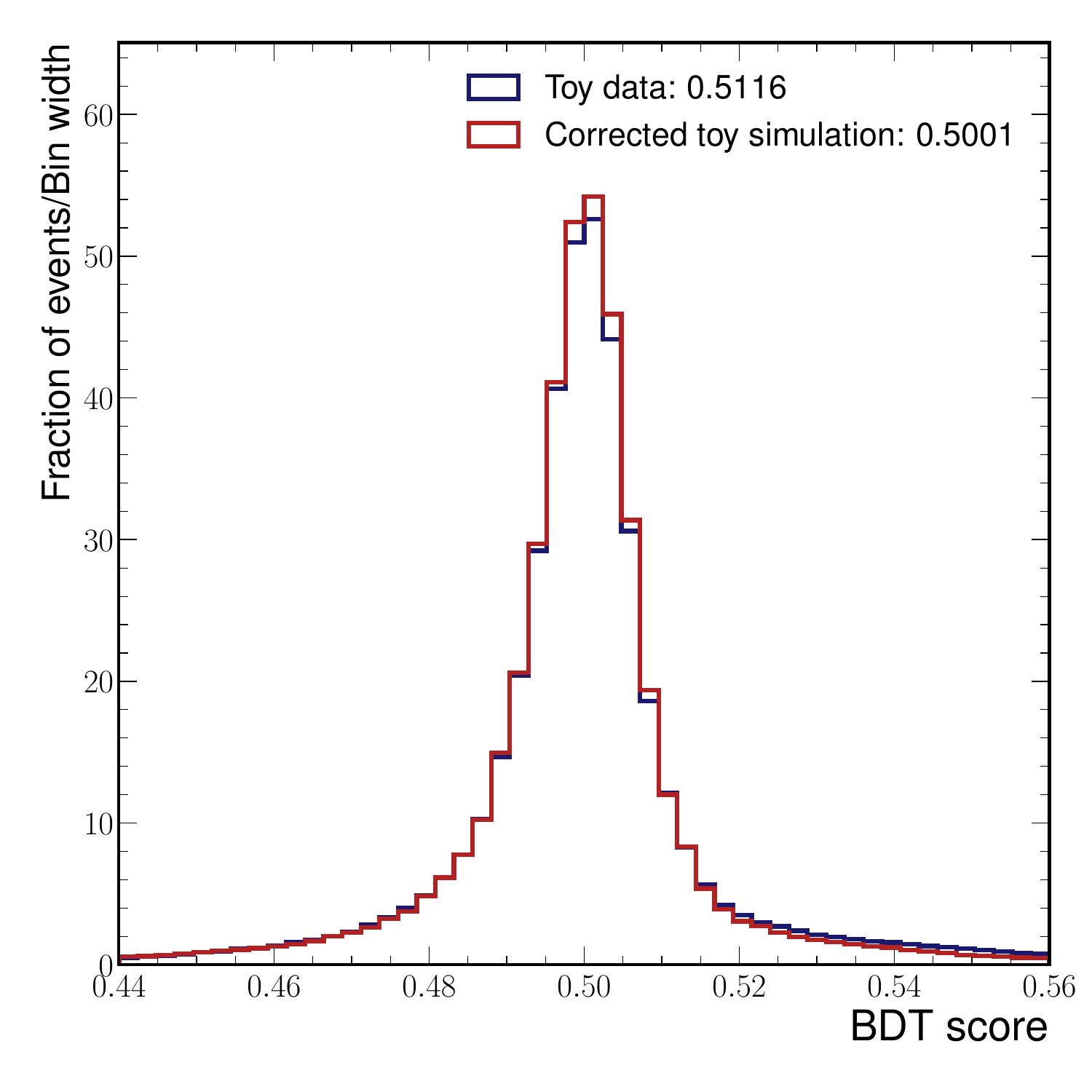}
        \caption{Output distributions of the BDT that was trained to distinguish data (blue) from corrected simulation (red), normalised to unit area.}
        \label{fig:BDToutput}
        \vspace{1cm}
        \includegraphics[width=0.6\textwidth]{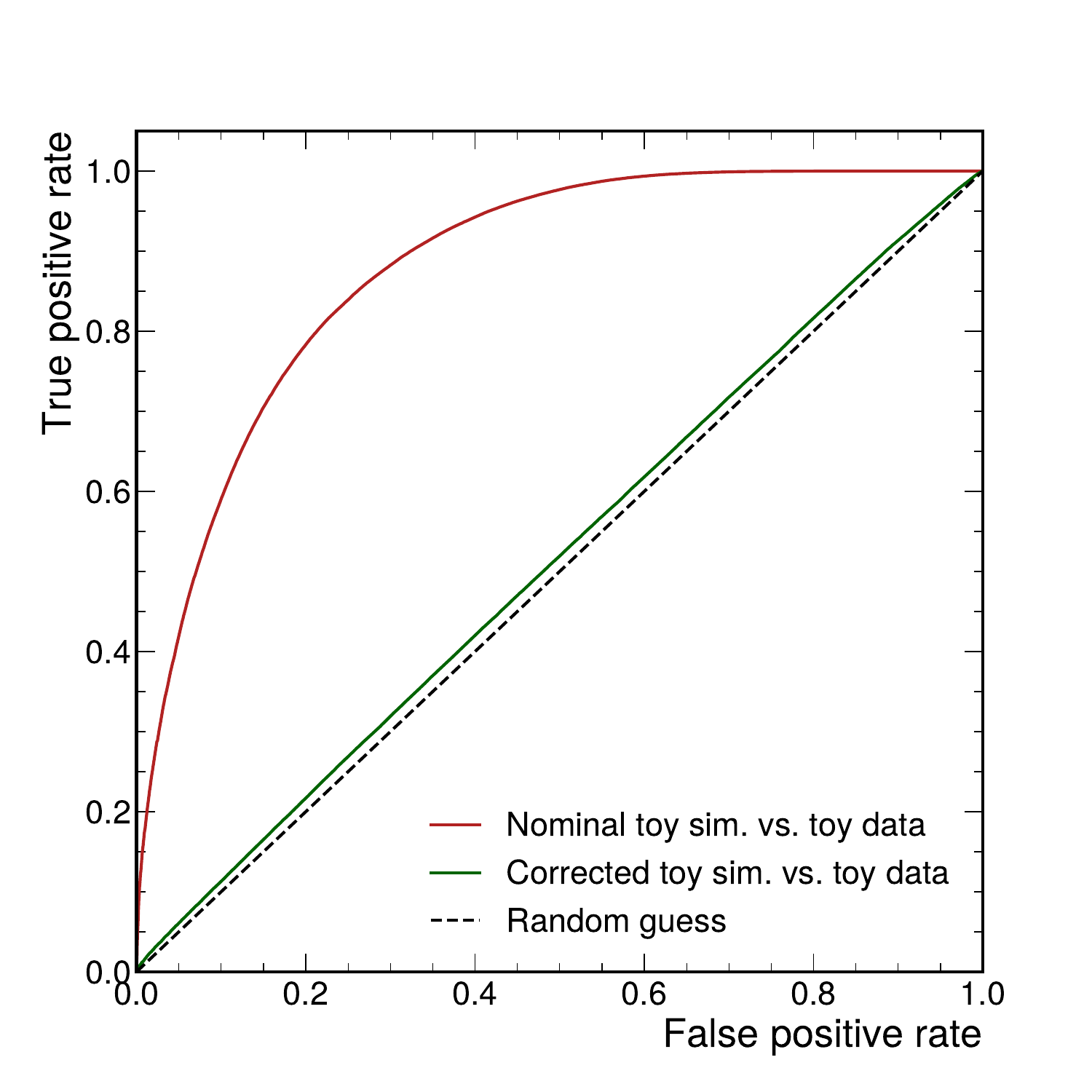}
        \caption{ROC curves for BDTs trained to separate between data and nominal (red) and corrected (green) simulation, evaluated on the test datasets (solid). The ROC curve for a random guess is shown as a black dashed line for reference. 
        }
        \label{fig:ROC}
\end{figure}

To assess whether the normalising-flow correction can also capture correlations correctly, we compare the Pearson correlation coefficients ($\rho$) between all seven variables (informative features and ancillary variables) for simulation and data.
In Figure~\ref{fig:corr_matrices}, we show the difference between the correlation coefficients between nominal simulation and data on the one hand and corrected simulation and data on the other hand.
The correlation coefficients for data and nominal simulation display notable differences, as induced during the generation of the dataset.
The comparison between data and corrected simulation reveals a significant improvement.
After the correction, the agreement in the correlation coefficients is below or at the level of $1\%$.
This shows that the normalising flow is able to learn and correct the non-trivial correlation structure in the datasets.

As a final assessment of the quality of the corrections, we train a BDT to distinguish corrected simulated samples from data.
We determine the classification power of the BDT to check how well the two underlying multivariate probability densities agree~\cite{lopezpaz2018revisiting}.
The idea is that the better the agreement between data and simulation, the more challenging it is for the BDT to differentiate between data samples and corrected simulation samples.
For comparison, we also train a separate BDT to distinguish data samples from uncorrected simulated samples.

The BDTs are trained with the \texttt{XGBoost} package~\cite{xgboost}, using logistic regression for binary classification as the loss function.
The training is stopped once the loss does not improve for 30~boosting rounds.
A learning rate of $0.1$ is used, and the trees can grow to a maximum depth of $10$.
For the training and evaluation of the BDTs, only the events from the original normalising-flow test dataset are used, which is again divided into a training ($60\%$) and test dataset ($40\%$).

The output distributions of the BDT that was trained to distinguish data and corrected simulation are shown in Figure~\ref{fig:BDToutput}.
The distributions for data and corrected simulation are very similar.
ROC curves for the two BDTs are shown in Figure~\ref{fig:ROC}.
The classifier is sufficiently powerful to distinguish data and uncorrected simulated events with an area under the curve (AUC) of $0.88$ on the test dataset.
Distinguishing data from corrected simulation is significantly more challenging: The ROC curve is close to the ROC curve for a random guess and has an AUC of $0.52$ on the test dataset.
These results indicate that the normalising-flow correction can also correct complex structural differences between the variables in the toy datasets that go beyond what can be captured with Pearson correlation coefficients.

We observed reduced BDT classification performance, i.e.~improved performance of the corrections, when we improved the reweighting of the ancillary variables in the preprocessing from uniform binning to binning with approximately the same number of events per bin.
This indicates that part of the residual difference between corrected simulation and data stems from small differences in the ancillary variables and not from imperfections in the normalising-flow morphing.

As an alternative to the autoregressive structure, we perform a test with a normalising flow based on GLOW coupling blocks~\cite{kingma2018glow}, where the affine transformation is replaced by the monotonic rational-quadratic transforms.
When training the BDT to discriminate simulations corrected with this alternative setup and data, we observe very similar performance with an AUC on the test dataset of~$0.52$.

\begin{figure}[ht]
    \centering
        \includegraphics[width=0.49\textwidth]{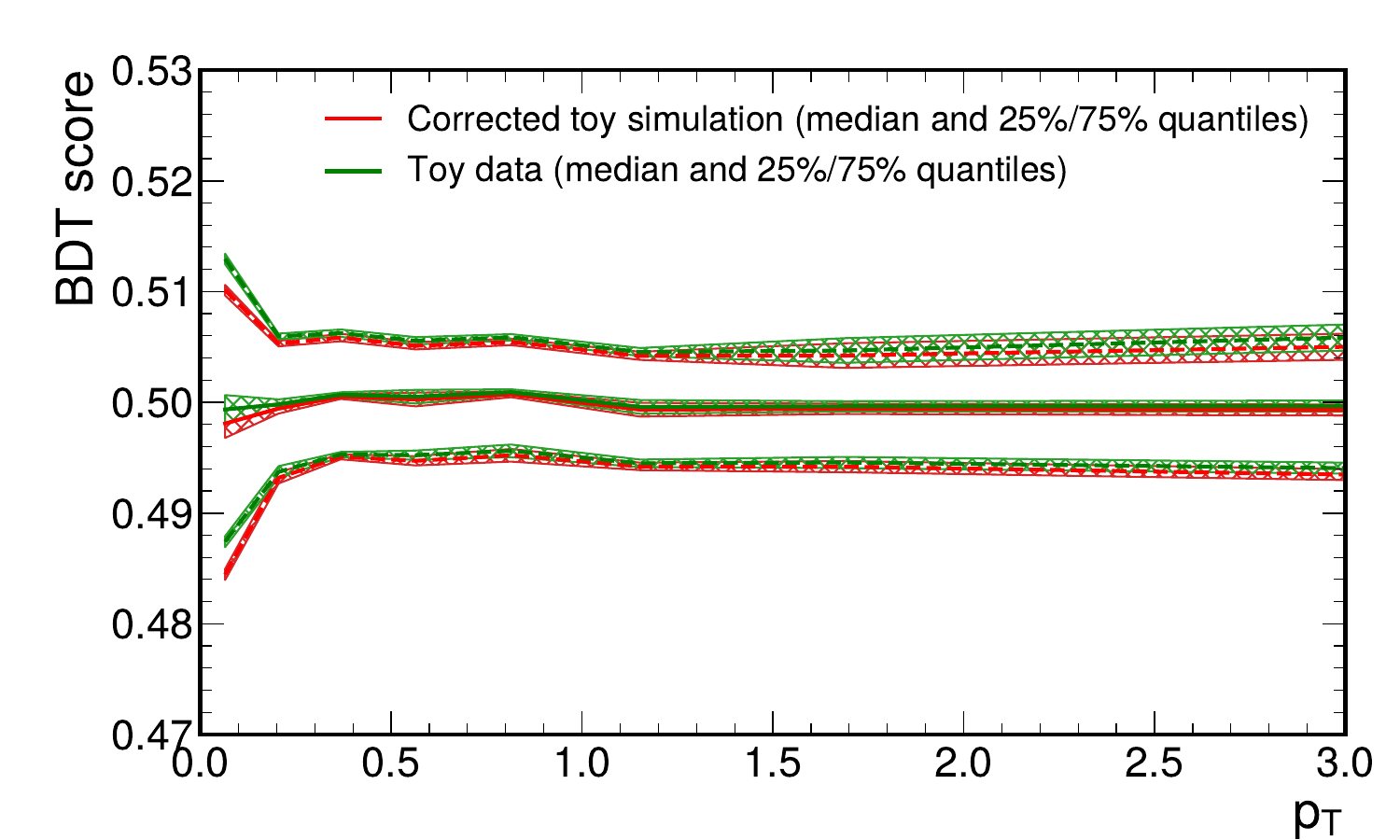}
        \includegraphics[width=0.49\textwidth]        {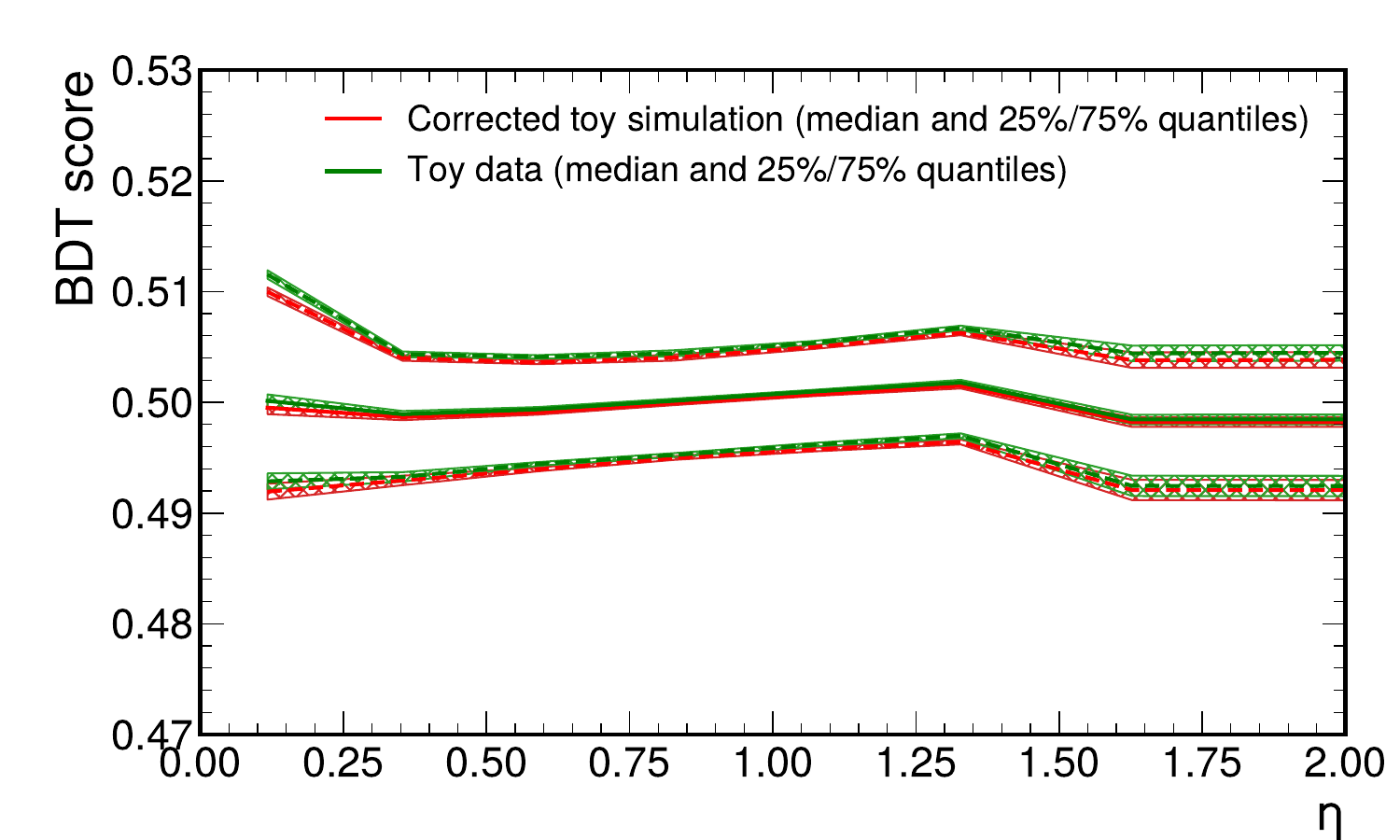}
        \includegraphics[width=0.49\textwidth]{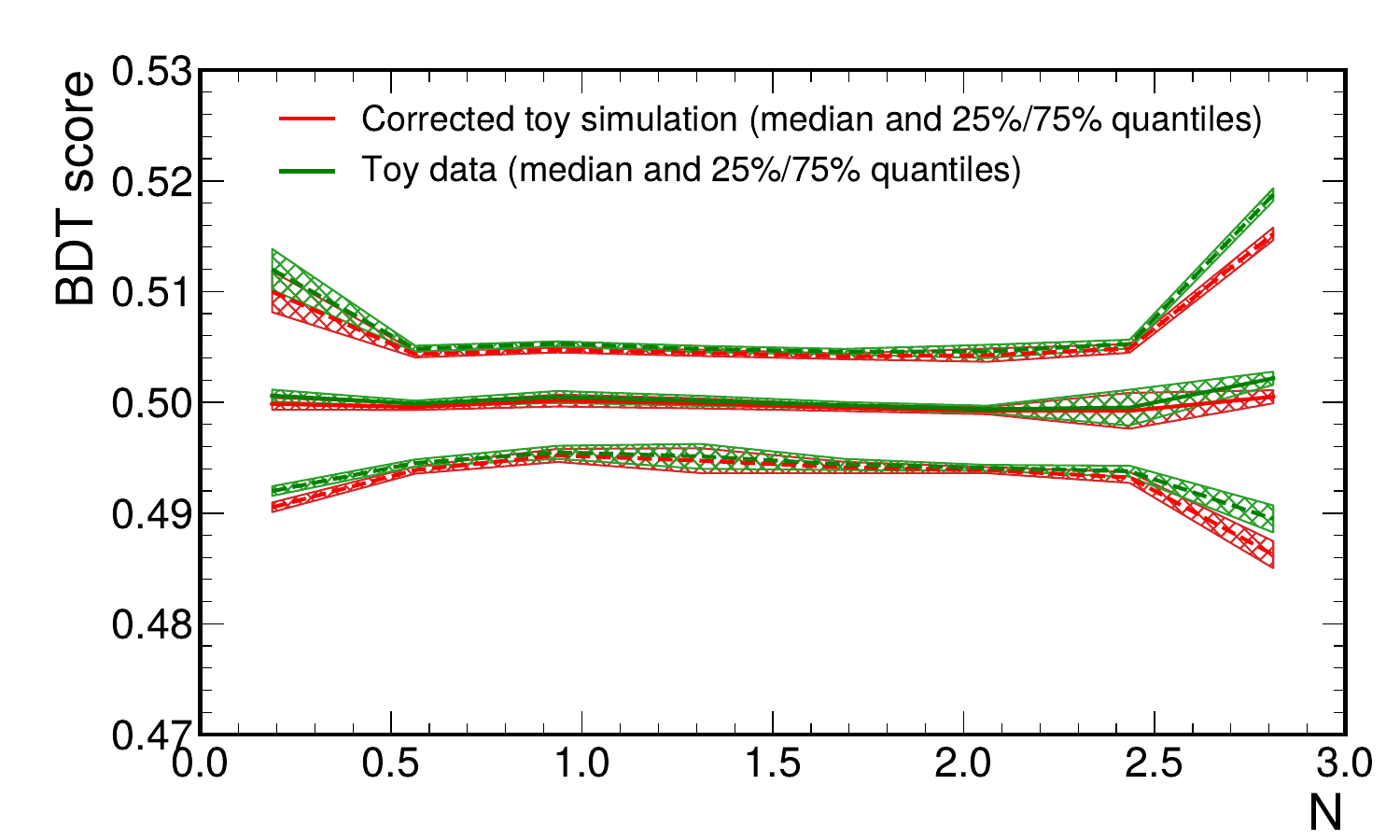}
    \caption{
    50\% (solid curves) and $25\%$ and $75\%$ quantiles (dashed curves) of the BDT output distribution as a function of \pt\ (top, left), \myEta\ (top, right) and \noise\ (bottom) for data (green) and corrected simulation (red).
    The quantiles are extracted in ten equally populated bins and the different values are shown at the bin centres.
    The ancillary variables are defined in a way that \pt\ is unitless, \myEta\ takes only positive values and \noise\ is defined in the interval $[0,3]$.
    Uncertainties from the limited number of simulated samples are shown as bands.}
    \label{fig:BDTscore_and_profiles}
\end{figure}

We also study the behaviour of the corrections over the full range of the ancillary features.
This is important for many applications in high-energy physics, in which the corrections are typically obtained from a phase space with a different distribution of the ancillary variables than the phase space where the corrections are ultimately applied.
We partition the events into 8 equally populated bins in \pt, \myEta\ and \noise.
For each bin, we obtain the $25\%$, 50\% and $75\%$ quantiles of the output distribution of the BDT that we trained to distinguish the data from the corrected simulation.
Figure~\ref{fig:BDTscore_and_profiles} shows these quantiles for data and corrected simulations as a function of \pt, \myEta and \noise.
The curves for data and corrected simulations are very similar, which indicates that the corrections are stable as a function of the three ancillary variables.

\FloatBarrier

\section{Conclusions}
\label{sec:conclusions}

We have introduced an approach for morphing two multidimensional distributions that is based on a single autoregressive normalising flow and a boolean variable as a condition.
We have applied this approach to correct simulation to correspond closer to data in a physics-inspired toy dataset of four informative features and three ancillary variables.
The toy dataset presents non-trivial correlations between variables and
pronounced differences in the distributions between simulation and data as a function of the ancillary variables.
In addition, we have described a smoothing procedure for discontinuous input distributions that is beneficial as a preprocessing step for normalising flows.

We find an agreement at the level of $1$-$2\%$ in the bulk of the marginal distributions of corrected simulation and data.
In the tails of the marginal distributions, corrected simulation and data agree at the level of a few percent.
In addition, we find large improvements in the agreement between the correlations in simulation and data.
Using a boosted decision tree to classify corrected simulation and data, we observe that both classes are hardly distinguishable after morphing and that the performance of the morphing is stable as a function of ancillary variables.

The morphing approach with a single normalising flow shows excellent performance and seems a promising tool for the correction of complex simulated distributions in high-energy physics and beyond.
Finally, we note that it is simple to extend the boolean switch to a multiclass condition to morph between more than two domains (cf. Appendix~\ref{app:3D}).

\subsubsection*{Acknowledgements}
The authors thank Nitish Kumar~K~V for cross-checking the generation of the toy datasets.
The authors acknowledge the support from the German Federal Ministry of Education and Research (BMBF) via project 05H2021 under grant number 05H23PACCA (CCD), the German Research Foundation (DFG) Heisenberg Programme under grant number 465126324 (JE), the DFG project 400140256 - GRK~2497 (JLS), the Studienstiftung des deutschen Volkes (JLS), and the Swiss National Science Foundation under contracts number 200020\_219622 and 200020\_200642 (MG, DV).

\clearpage

\appendix

\section{Two-dimensional visualisation of the dataset}
\label{app:cornerplot}

\begin{figure}[hbt!] 
    \centering
    \includegraphics[width=\linewidth]{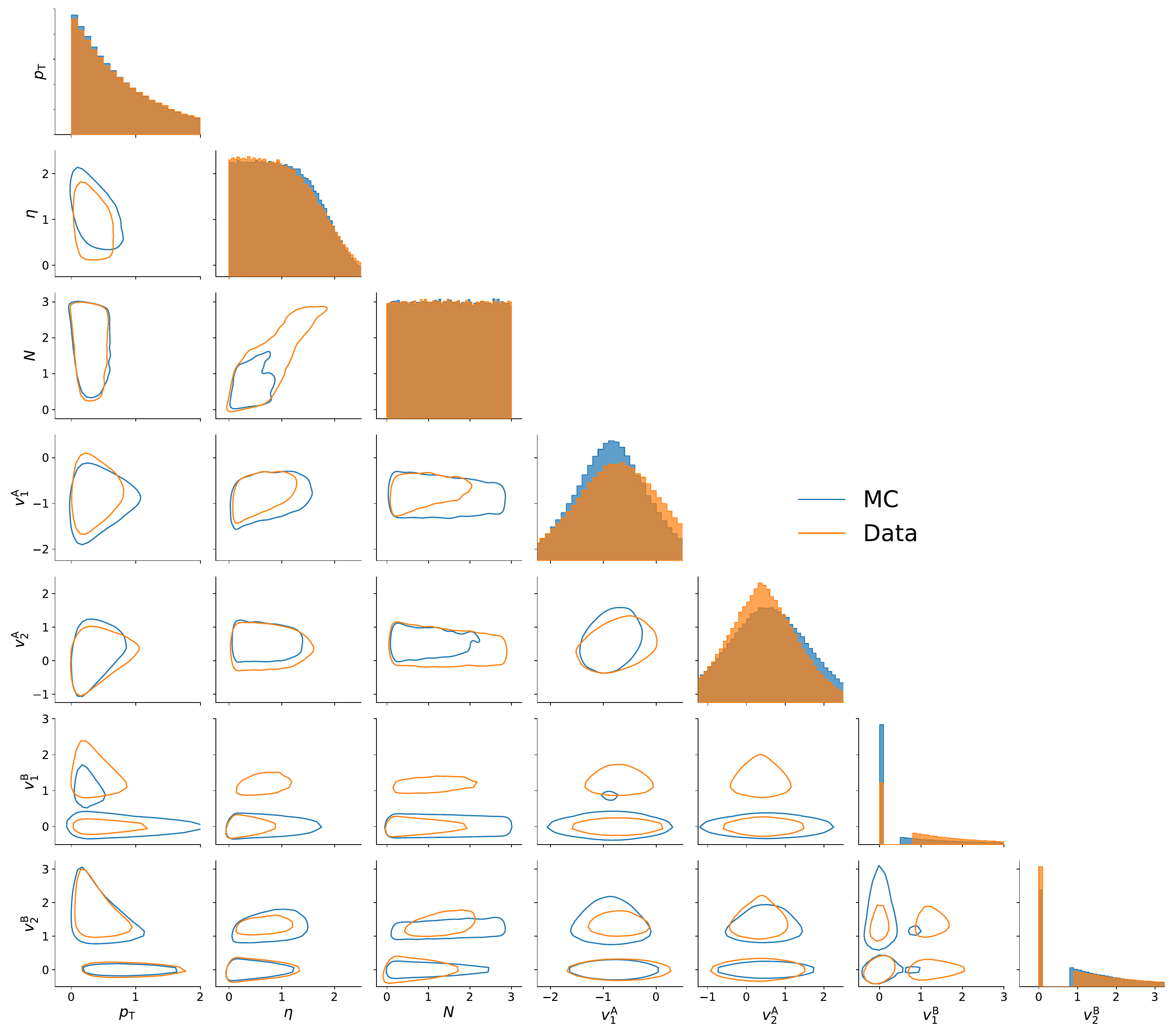}
    \caption{
    \textcolor{black}{Visualisation of the physics-inspired dataset in selected ranges for simulation (blue) and data (orange). 
    On the diagonal, the marginal distributions are shown.
    On the off-diagonal, the $68\%$ contours obtained from kernel density esimation are depicted.
    The ancillary variables are defined in a way that \pt\ is unitless, \myEta\ takes only positive values and \noise\ is defined in the interval [0,3].}}
    \label{fig:cornerplot}
\end{figure}

\textcolor{black}{Figure~\ref{fig:cornerplot} visualises the physics-inspired dataset by showing two-dimensional 68\% contours of each pair of variables.
This shows their non-trivial correlations and highlights that these are different between toy simulation and toy data.}

\section{Details of the dataset generation}
\label{app:gen_formulas}

The \pt\ variable is distributed according to the density
\begin{equation}
    f(\pt;\beta_p) = \frac{1}{\beta_p}\exp(-\pt/\beta_p)
\end{equation}
with scale parameter $\beta_p$.
The values for the \myEta\ variable are generated according to $\myEta = \lvert u \cdot n \rvert$\,, where the two factors are distributed according to a uniform and a Gaussian distribution, respectively: $U \sim \mathcal{U}(-2,2)$ and $N \sim \mathcal{N}(1, \eta_\mathrm{s})$.
The noise \noise\ is uniformly distributed in the interval $[0,3]$.

The values for the \viA variables are drawn from uncorrelated Gaussian distributions with conditioned means of $\mu - f_\mu \cdot \myEta$ and standard deviations of $\mathrm{max}(\sigma - f_\sigma \cdot \pt\ + 0.1 \cdot \noise,0.1)$.
The \viB values are drawn from a mixture distribution composed of a Dirac delta peak at zero and a shifted exponential distribution.
Its density has five parameters: $r$, $f_r$, $x_\mathrm{t}$, $\beta_\mathrm{B}$, and $f_\mathrm{B}$.
The probability of assigning a value of zero is given by $r + f_r \cdot \pt - 0.1 \cdot \noise$.
The exponential distributions is shifted by $x_\mathrm{t}$ to the right and the scale parameter is given by $\beta_\mathrm{B} + f_\mathrm{B} \cdot \myEta + 0.2 \cdot \noise.$
The parameter values in the \viA\ and \viB\ families differ for the individual variables.

The specific values for the parameters that were used to generate the samples for this study are given in Table~\ref{tab:settings}.

\begin{table}[htbp]
\centering
\begin{tabular}{c|c|c|c c c c|c c c c c}
& $\beta_p$ & $\eta_\mathrm{s}$ & $\mu$ & $f_\mu$ & $\sigma$ & $f_\sigma$ & $r$ & $f_r$ & $x_\mathrm{t}$ & $\beta_\mathrm{B}$ & $f_\mathrm{B}$ \\
\hline
\hline
\pt & $1.0$ & & & & & & & & & & \\
\myEta & & $0.25$ & & & & & & & & & \\
\vOneA & & & $-0.5$ & $0.2$ & $0.8$ & $0.1$ & & & & & \\
\vTwoA & & & $0.5$ & $0.1$ & $1.0$ & $0.2$ & & & & & \\
\vOneB & & & & & & & $0.3$ & $0.1$ & $0.8$ & $1.0$ & $0.3$ \\
\vTwoB & & & & & & & $0.4$ & $0.05$ & $0.9$ & $1.3$ & $0.2$ \\
\hline
\hline
\pt & $0.95$ & & & & & & & & & & \\
\myEta & & $0.2$ & & & & & & & & & \\
\vOneA & & & $-0.6$ & $0.25$ & $0.7$ & $0.15$ & & & & & \\
\vTwoA & & & $0.6$ & $0.05$ & $1.1$ & $0.15$ & & & & & \\
\vOneB & & & & & & & $0.5$ & $0.15$ & $0.5$ & $1.1$ & $0.35$ \\
\vTwoB & & & & & & & $0.3$ & $0.1$ & $0.8$ & $1.1$ & $0.15$ \\
\end{tabular}
\caption{Parameter settings for the generation of the toy dataset used in this study.
The values in the upper (lower) part were used to generate toy data (simulation).}
\label{tab:settings}
\end{table}

After the generation of these uncorrelated variables, the \texttt{induce\_correlations} function of the \texttt{mcerp} package~\cite{mcerp, mcerp_theory} is used to bestow non-trivial correlations among all seven variables.
The degenerate values at zero for the \viB variables are smeared with small uniform noise to avoid that the correlation algorithm fails, as it is based on the per-variable ranking of the input data.
The smeared values are mapped back to zero after the correlations are induced.

\section{Morphing between three domains}
\label{app:3D}

\begin{figure}[!ht]
    \centering
        \centering
        \includegraphics[width=\textwidth]{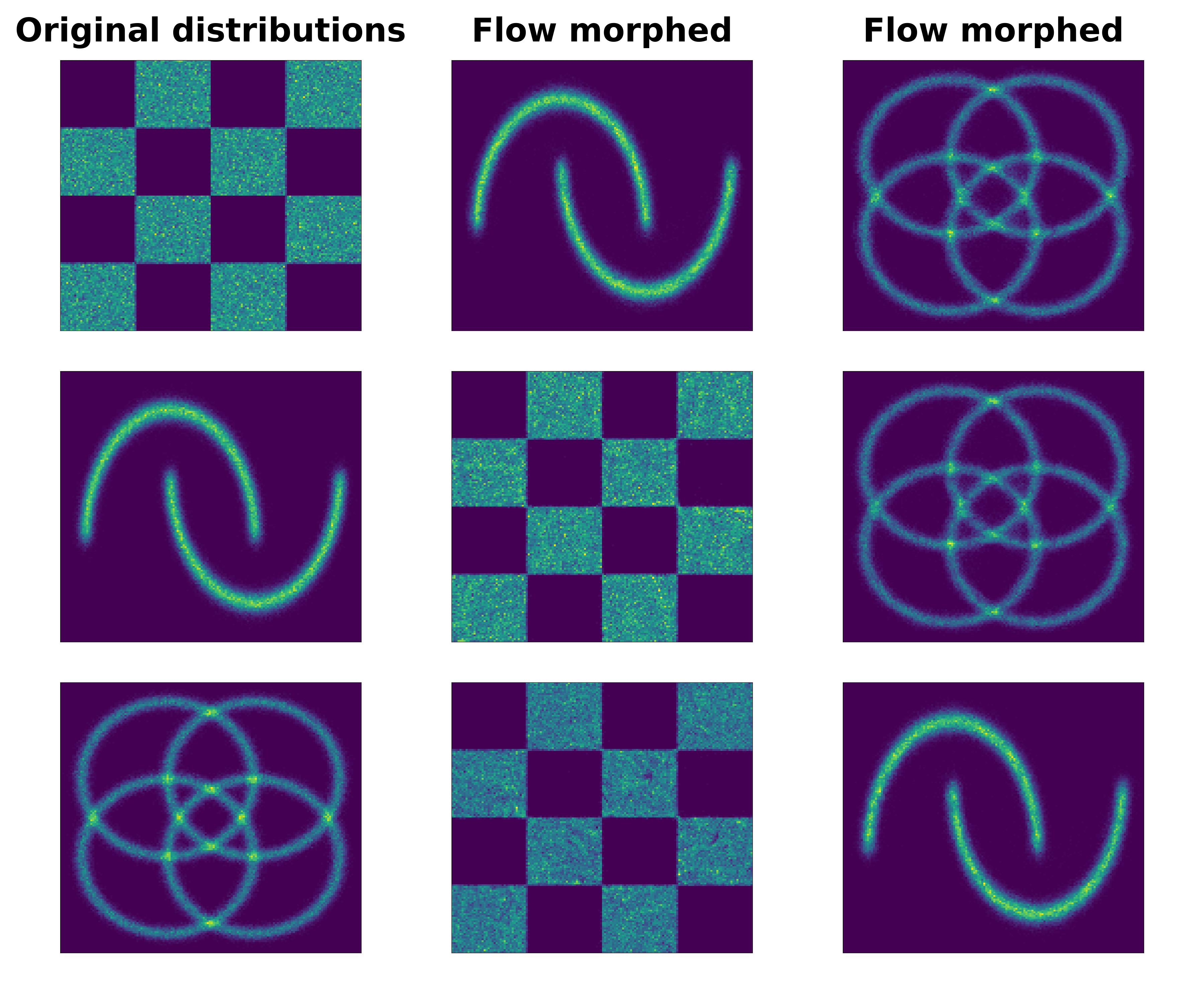}
        \caption{Morphing between three domains: The upper row shows the morphing from the checkerboard to the two-moons and four-circles datasets. The middle (bottom) row shows the morphing from the two-moons (four-circles) to the other two datasets.
        }
        \label{fig:Three_distributions}
\end{figure}

The single-flow morphing can easily be extended to morph between more than two domains by using one-hot encoding instead of a boolean condition.
We illustrate this in Figure~\ref{fig:Three_distributions} by showing the results of morphing between three two-dimensional distributions: checkerboard, two moons and four circles.
As in the two-domain case in Section~\ref{sec:2D}, the single-flow approach is able to reproduce the sharp edges and discontinuous features of the distributions.

\FloatBarrier

\phantomsection
\addcontentsline{toc}{section}{References}
\bibliographystyle{JHEP_jls}
\bibliography{paper_Nflows.bib}

\providecommand{\href}[2]{#2}\begingroup\raggedright\begin{thebibliography}{10}

\bibitem{Cranmer:2015bka}
K.~Cranmer, J.~Pavez and G.~Louppe, \emph{{Approximating Likelihood Ratios with Calibrated Discriminative Classifiers}},  \href{https://arxiv.org/abs/1506.02169}{{\ttfamily 1506.02169}}.

\bibitem{Rogozhnikov:2016bdp}
A.~Rogozhnikov, \emph{{Reweighting with Boosted Decision Trees}}, \href{http://dx.doi.org/10.1088/1742-6596/762/1/012036}{\emph{J. Phys. Conf. Ser.} {\bfseries 762} (2016) 012036}, [\href{https://arxiv.org/abs/1608.05806}{{\ttfamily 1608.05806}}].

\bibitem{Andreassen:2019nnm}
A.~Andreassen and B.~Nachman, \emph{{Neural Networks for Full Phase-space Reweighting and Parameter Tuning}}, \href{http://dx.doi.org/10.1103/PhysRevD.101.091901}{\emph{Phys. Rev. D} {\bfseries 101} (2020) 091901}, [\href{https://arxiv.org/abs/1907.08209}{{\ttfamily 1907.08209}}].

\bibitem{Diefenbacher_2020_DCTRGAN}
S.~Diefenbacher, E.~Eren, G.~Kasieczka, A.~Korol, B.~Nachman and D.~Shih, \emph{{DCTRGAN: improving the precision of generative models with reweighting}}, \href{http://dx.doi.org/10.1088/1748-0221/15/11/P11004}{\emph{JINST} {\bfseries 15} (2020) P11004}.

\bibitem{CQR_Higgs_full_run2}
{\scshape CMS} collaboration, A.~Tumasyan et~al., \emph{{Measurement of the Higgs boson inclusive and differential fiducial production cross sections in the diphoton decay channel with pp collisions at $ \sqrt{s} $ = 13 TeV}}, \href{http://dx.doi.org/10.1007/JHEP07(2023)091}{\emph{JHEP} {\bfseries 07} (2023) 091}, [\href{https://arxiv.org/abs/2208.12279}{{\ttfamily 2208.12279}}].

\bibitem{Erdmann:2018kuh}
M.~Erdmann, L.~Geiger, J.~Glombitza and D.~Schmidt, \emph{{Generating and refining particle detector simulations using the Wasserstein distance in adversarial networks}}, \href{http://dx.doi.org/10.1007/s41781-018-0008-x}{\emph{Comput. Softw. Big Sci.} {\bfseries 2} (2018) 4}, [\href{https://arxiv.org/abs/1802.03325}{{\ttfamily 1802.03325}}].

\bibitem{amos2017input}
B.~Amos, L.~Xu and J.~Z. Kolter, \emph{Input convex neural networks},  \href{https://arxiv.org/abs/1609.07152}{{\ttfamily 1609.07152}}.

\bibitem{Pollard:2021fqv}
C.~Pollard and P.~Windischhofer, \emph{{Transport away your problems: Calibrating stochastic simulations with optimal transport}}, \href{http://dx.doi.org/10.1016/j.nima.2021.166119}{\emph{Nucl. Instrum. Meth. A} {\bfseries 1027} (2022) 166119}, [\href{https://arxiv.org/abs/2107.08648}{{\ttfamily 2107.08648}}].

\bibitem{Butter:2023ira}
A.~Butter, T.~Jezo, M.~Klasen, M.~Kuschick, S.~Palacios~Schweitzer and T.~Plehn, \emph{{Kicking it Off(-shell) with Direct Diffusion}},  \href{https://arxiv.org/abs/2311.17175}{{\ttfamily 2311.17175}}.

\bibitem{Algren:2023qnb}
M.~Algren, T.~Golling, M.~Guth, C.~Pollard and J.~A. Raine, \emph{{Flow Away your Differences: Conditional Normalizing Flows as an Improvement to Reweighting}},  \href{https://arxiv.org/abs/2304.14963}{{\ttfamily 2304.14963}}.

\bibitem{CQRwithFlows}
S.~Bright-Thonney, P.~Harris, P.~McCormack and S.~Rothman, \emph{{Chained Quantile Morphing with Normalizing Flows}},  \href{https://arxiv.org/abs/2309.15912}{{\ttfamily 2309.15912}}.

\bibitem{flow4flows}
T.~Golling, S.~Klein, R.~Mastandrea, B.~Nachman and J.~A. Raine, \emph{{Morphing one dataset into another with maximum likelihood estimation}}, \href{http://dx.doi.org/10.1103/PhysRevD.108.096018}{\emph{Phys. Rev. D} {\bfseries 108} (2023) 096018}, [\href{https://arxiv.org/abs/2309.06472}{{\ttfamily 2309.06472}}].

\bibitem{Tabak:2013cnz}
E.~G. Tabak and C.~V. Turner, \emph{{A Family of Nonparametric Density Estimation Algorithms}}, \href{http://dx.doi.org/10.1002/cpa.21423}{\emph{Commun. Pure Appl. Math.} {\bfseries 66} (2013) 145--164}.

\bibitem{papamakarios2021normalizing_review}
G.~Papamakarios, E.~Nalisnick, D.~J. Rezende, S.~Mohamed and B.~Lakshminarayanan, \emph{Normalizing flows for probabilistic modeling and inference}, \href{http://dx.doi.org/https://doi.org/10.48550/arXiv.1912.02762}{\emph{J. Mach. Learn. Res.} {\bfseries 22} (2021) 1--64}, [\href{https://arxiv.org/abs/1912.02762}{{\ttfamily 1912.02762}}].

\bibitem{Kobyzev_2021}
I.~Kobyzev, S.~J. Prince and M.~A. Brubaker, \emph{Normalizing flows: An introduction and review of current methods},  \href{https://arxiv.org/abs/1908.09257}{{\ttfamily 1908.09257}}.

\bibitem{papamakarios2018masked_MAF}
G.~Papamakarios, T.~Pavlakou and I.~Murray, \emph{Masked autoregressive flow for density estimation},  \href{https://arxiv.org/abs/1705.07057}{{\ttfamily 1705.07057}}.

\bibitem{müller2019neural_linear_quadratic_splines}
T.~Müller, B.~McWilliams, F.~Rousselle, M.~Gross and J.~Novák, \emph{Neural importance sampling},  \href{https://arxiv.org/abs/1808.03856}{{\ttfamily 1808.03856}}.

\bibitem{durkan2019cubicspline}
C.~Durkan, A.~Bekasov, I.~Murray and G.~Papamakarios, \emph{Cubic-spline flows},  \href{https://arxiv.org/abs/1906.02145}{{\ttfamily 1906.02145}}.

\bibitem{linear_rational_spline}
C.~L. Hadi M.~Dolatabadi, Sarah~Erfani, \emph{Invertible generative modeling using linear rational splines},  \href{https://arxiv.org/abs/2001.05168}{{\ttfamily 2001.05168}}.

\bibitem{Spline_flows}
C.~Durkan, A.~Bekasov, I.~Murray and G.~Papamakarios, \emph{Neural spline flows},  \href{https://arxiv.org/abs/1906.04032}{{\ttfamily 1906.04032}}.

\bibitem{germain2015made}
M.~Germain, K.~Gregor, I.~Murray and H.~Larochelle, \emph{Made: Masked autoencoder for distribution estimation},  \href{https://arxiv.org/abs/1502.03509}{{\ttfamily 1502.03509}}.

\bibitem{sklearn}
F.~Pedregosa, G.~Varoquaux, A.~Gramfort, V.~Michel, B.~Thirion, O.~Grisel et~al., \emph{Scikit-learn: Machine learning in {P}ython}, {\emph{J. Mach. Learn. Res.} {\bfseries 12} (2011) 2825--2830}, [\href{https://arxiv.org/abs/1201.0490}{{\ttfamily 1201.0490}}].

\bibitem{zuko}
F.~Rozet et~al., \emph{{Zuko}: Normalizing flows in pytorch},  \url{https://pypi.org/project/zuko}.
\newblock 10.5281/zenodo.7625672.

\bibitem{pytorch}
A.~Paszke, S.~Gross, F.~Massa, A.~Lerer, J.~Bradbury, G.~Chanan et~al., \emph{Pytorch: An imperative style, high-performance deep learning library},  \href{https://arxiv.org/abs/1912.01703}{{\ttfamily 1912.01703}}.

\bibitem{Adam_optimizer}
D.~P. Kingma and J.~Ba, \emph{Adam: A method for stochastic optimization},  \href{https://arxiv.org/abs/1412.6980}{{\ttfamily 1412.6980}}.

\bibitem{cossine_annealing}
I.~Loshchilov and F.~Hutter, \emph{Sgdr: Stochastic gradient descent with warm restarts},  \href{https://arxiv.org/abs/1608.03983}{{\ttfamily 1608.03983}}.

\bibitem{mcerp}
A.~Lee, \emph{Mcerp: Monte carlo error propagation},  \url{https://pythonhosted.org/mcerp/}.

\bibitem{mcerp_theory}
R.~L. Iman and W.~J. Conover, \emph{A distribution-free approach to inducing rank correlation among input variables}, \href{http://dx.doi.org/10.1080/03610918208812265}{\emph{Commun. Stat. Simul. Comput.} {\bfseries 11} (1982) 311--334}.

\bibitem{PhysRevD.107.113003_CALOflow}
C.~Krause and D.~Shih, \emph{Fast and accurate simulations of calorimeter showers with normalizing flows}, \href{http://dx.doi.org/10.1103/PhysRevD.107.113003}{\emph{Phys. Rev. D} {\bfseries 107} (2023) 113003}.

\bibitem{lopezpaz2018revisiting}
D.~Lopez-Paz and M.~Oquab, \emph{Revisiting classifier two-sample tests},  \href{https://arxiv.org/abs/1610.06545}{{\ttfamily 1610.06545}}.

\bibitem{xgboost}
T.~Chen and C.~Guestrin, \emph{Xgboost: A scalable tree boosting system},  \href{https://arxiv.org/abs/1603.02754}{{\ttfamily 1603.02754}}.

\bibitem{kingma2018glow}
D.~P. Kingma and P.~Dhariwal, \emph{Glow: Generative flow with invertible 1x1 convolutions},  \href{https://arxiv.org/abs/1807.03039}{{\ttfamily 1807.03039}}.

\end{thebibliography}\endgroup

\end{document}